\documentclass[probth,final]{sw}
\usepackage{latexsym}
\usepackage{amsmath,amsfonts,amssymb}
\usepackage{graphics}
\usepackage{times}

\numberwithin{theorem}{section}
\numberwithin{proposition}{section}
\numberwithin{corollary}{section}
\numberwithin{lemma}{section}
\numberwithin{definition}{section}
\numberwithin{remark}{section}
\numberwithin{equation}{section}

\spnewtheorem{remarks}{Remarks}{\bf}{\rm}
\numberwithin{remarks}{section}
\def\C{\mathbb C}
\def\Z{\mathbb Z}

\def\T{\mathbb T}
\def\N{\mathbb N}
\def\R{\mathbb R}
\def\P{\mathbb P}

\def\tr{\operatorname{Tr}}    

\makeatletter
\let\Im\undefined
\let\Re\undefined
\DeclareMathOperator{\Im}{Im}
\DeclareMathOperator{\Re}{Re}

\newcounter{numcount}
\newcommand{\labelnummer}{\mbox{(\roman{numcount})}}%

\newenvironment{indentnummer*}%
      {\begin{list}{\labelnummer}{\usecounter{numcount}%
                    \topsep1.2ex\partopsep2ex\parsep0pt\itemsep1ex
                    \labelwidth2.5em\itemindent0em\labelsep1em%
                    \leftmargin3.5em
                    }}%
     {\end{list}}%

\makeatother
\begin{document}
\title{Stability of the Absolutely Continuous Spectrum of Random
Schr\"odinger Operators on Tree Graphs}

\titlerunning{Stability of the Absolutely Continuous Spectrum}
\authorrunning{Aizenman, Sims, and Warzel}
\author{Michael Aizenman $^{\rm (1,a)}$
   \thanks{$^{\rm a}$ Visiting the Department of Physics
    of Complex System, Weizmann Inst. of Science, Israel \\
$^{\rm b}$ Present address: 
Department of Mathematics, University of California at Davis, 
Davis CA 95616, USA\\
$^{\rm c}$
           On leave from: Institut f\"ur Theoretische Physik,
Universit\"at Erlangen-N\"urnberg, Germany}
\and Robert Sims $^{\rm (1,b)}$
\and  Simone Warzel $^{\rm (1,c)}$ 
}                     
%
%
\institute{$^1$ Departments of Physics and Mathematics,
      Princeton University, Princeton NJ 08544, USA
}
\date{Received: February 1, 2005 / Revised: June 23, 2005}
%
\maketitle
\begin{abstract}
%
The subject of this work is random Schr\"odinger operators 
on regular rooted tree graphs $\T$ with stochastically homogeneous disorder.  
The operators are of 
the form   $ H_\lambda(\omega) = T + U + \lambda V(\omega) $ acting in
$\ell^2(\T)$, with  
$T $ the adjacency  matrix, $U$ a radially periodic potential, and $V(\omega)$ 
 a  random potential.     
This includes the only class of homogeneously random operators
for which it was proven  that  the spectrum of  $H_\lambda(\omega)$
exhibits an absolutely continuous (ac) component; a results established by
A. Klein for weak disorder in case $U=0$ and $V(\omega)$ 
given by  iid random variables on $\T$.
Our main contribution is a new method for establishing the persistence of
ac spectrum under weak disorder.   The method yields 
the continuity of  the ac 
spectral density of $H_\lambda(\omega)$ at $\lambda = 0$.  
The latter  is shown to converge  in the $L^1$-sense over closed Borel sets in 
which $H_0$ has no singular spectrum.  The analysis extends  to random potentials 
whose values at different
sites need not be independent, assuming only that their joint  
distribution  is  weakly  correlated across different tree branches.
\end{abstract}
\setcounter{tocdepth}{2}
\tableofcontents

\section{Introduction}

The objective of this work is to present results on the stability of the
absolutely continuous spectrum of Schr\"odinger operators on
tree graphs, under  the addition of weak but extensive disorder in
the form of a random potential.

The background for this analysis is  the generally known
phenomenon of {\em Anderson localization}: the addition of extensive
disorder  to a linear operator results in the localization of the
eigenfunctions  corresponding to certain spectral regimes, where the
spectral type changes to {\em pure-point}.  The localization regime
may  cover the full spectral range -- as is typically the case in one
dimension even at arbitrarily small, but non-zero  strength of the
disorder \cite{Anderson,GoMo77,CaLa90,PF,Stoll}.  
A major challenge for analysts is to shed light on extended states 
and ac spectrum.   We shall not review here the 
growing body of interesting works on decaying disorder, as our main 
focus concerns the homogeneous case.    
In this case the only proof of the  persistence of 
{\em de-localization}   -- in the sense
of the existence of extended states, or of absolutely continuous (ac)
spectrum in a certain energy range -- was obtained for the 
Laplacian on a regular tree perturbed by a weak random
potential  which is given by a collection  of iid 
 random variables~\cite{Klein95,Klein98}.   
In this work we return to the tree
setup and present a different set of tools.

\subsection{Statement of the main result}\label{sec:main}

We consider random Schr\"odinger operators on the Hilbert space
$\ell^2(\T)$ where $\T$ is
the set of vertices of a regular rooted tree graph 
in which each 
vertex has $K\ge 2$ forward neighbors 
(see Subsection~\ref{Sec:Ass}
for some of the basic terminology).
These operators are linear and of the form
\begin{equation} \label{eq:H}
H_{\lambda}(\omega) :=   T + U   + \lambda \, V(\omega)
\end{equation}
where:
\begin{enumerate}
\item   The operator $T$  corresponds to the adjacency matrix, i.e.,
the discrete version of the Laplacian
          without the diagonal terms:
          \begin{equation}
           \big(T \psi\big)_x := \sum_{ y}\, \psi_y \qquad
          \mbox{for all $ \psi \in \ell^2\big(\mathbb{T}\big)$,}
\end{equation}
where the sum runs over all nearest neighbor vertices of $ x \in \mathbb{T}$.
\item The term $U$ is a multiplication operator by a real-valued function $ \{U_x \}_{x \in \T} $ which is
          {radial} and $\tau$-{periodic} in $ | x | $, the distance  to the root.
\item   The real parameter $\lambda$ controls the strength of the
random perturbation.
\item  The symbol $\omega  $ represents the
randomness, i.e., $V(\omega)  $
          is a multiplication operator which is given in terms of an element $ \{ \omega_x \}_{x \in \T } $
        from the probability space~$\big(\R^\T,\P\big)$.
          Averages over that probability space will be denoted below by
${\mathbb E[ \,\cdot\,]} $.
\end{enumerate}
For each $\lambda$ and $ \omega $ the operator $H_{\lambda}(\omega) $
is essentially self adjoint on the domain of functions of compact
support. It is important for our discussion that the unperturbed part
\begin{equation}
H_0 = T+ U
\end{equation}
is a radially periodic Schr\"odinger operator on $\ell^2(\T)$
in the sense described in 1. and 2. above.
In order to prepare for the statement of our main result, let us note two facts
about the spectra of such operators (see~Appendix~\ref{app:proof}).

\begin{proposition}\label{prop:H0}
Let $ U $ be radial and periodic and $ V(\omega) $ be
\emph{radial}, i.e., $
          V_x(\omega) = \omega_{|x|} $, with $\{ \omega_{n} \}_{n \in \mathbb{N}_0}$
iid \emph{non-constant} random variables.
Then:
\begin{enumerate}\itemsep0.5ex
\item The ac spectrum of $H_0 = T + U $ on $ \ell^2(\T) $ consists of
a finite union of closed intervals. 
\item  For every $\lambda \neq 
0$ the ac spectrum of $ 
H_\lambda(\omega)= H_0 + \lambda
V(\omega) $ vanishes for almost all $ \omega $.
\end{enumerate}
\end{proposition}

Our main result is that the effect on the ac spectrum is different
when the perturbation is by a random potential
whose values over different branches of the tree are only weakly correlated.
Altogether, in this paper the random potential~$ V(\omega) $ is assumed to have
the following properties, whose
precise definitions can be found in Subsection~\ref{Sec:Ass} below.
\begin{indentnummer*}
\item[{\bf A1:}]  The probability measure $ \P $ of the random potential
is stationary under the symmetries associated with the graph 
endomorphisms of the rooted tree.
\item[{\bf A2:}] The values of the potential are log-integrable:  $
\mathbb{E} \big[ \log (1+|V_{x}(\cdot) |) \big]  < \infty  $ for each
$x\in\T$.
\item[{\bf A3:}] The probability measure $ \P $ of the random
potential is weakly correlated.
\end{indentnummer*}
These assumptions in particular ensure that the ac spectrum of $ H_\lambda(\omega) $ coincides with a non-random Borel set 
$ \Sigma_{\rm ac}(\lambda) $ for almost all $ \omega $.\\ 

Following is our main result.
\begin{theorem}\label{thm:main}
Let $ U $ be radial and periodic and
$V(\omega) $ satisfy {\bf A1}, {\bf A2} and {\bf A3}.
Then the random Schr\"odinger operator
$H_{\lambda}(\omega)=H_0+\lambda V(\omega)$ has the following
properties.    
\begin{enumerate}\itemsep0.5ex
\item    The ac spectrum is continuous at $ \lambda = 0 $ in the sense that for any Borel set 
        $I\subseteq \Sigma_{\rm ac} (0) $
        \begin{equation} \label{eq:main1a}
        \lim_{\lambda \to 0 } \; \mathcal{L}\left[ I \cap \Sigma_{\rm  ac}(\lambda) \right] =  
        \mathcal{L}\left[ I \cap \Sigma_{\rm  ac}(0) \right] 
\end{equation}
where $ \mathcal{L} $ denotes the Lebesgue measure.
\item  Over closed Borel sets $I\subseteq \Sigma_{\rm ac} (0) $ 
which 
are  free of singular spectrum of $H_0$ 
the density of the ac 
component of the
    spectral measure associated with $ \delta_0 $ is
    $L^1$-continuous at $\lambda = 0$ in the sense that:
\begin{multline}\label{eq:main1b}
\lim_{\lambda \to 0} \int_I \mathbb{E}\left[ \left|\, \Im \,\langle
\delta_0, (H_{\lambda}(\cdot)-E-i0)^{-1}  \delta_0\rangle  \right.\right.\\
        \left.\left.    -  \, \Im \, \langle \delta_0,
(H_0-E-i0)^{-1}  \delta_0\rangle \right| \right] \, dE  = 0.
\end{multline}
Here $\delta_0 \in \ell^2(\T)$ is the indicator function supported at the root.
\end{enumerate}
\end{theorem}

\subsection{The assumptions}\label{Sec:Ass}

A rooted tree is a connected, undirected graph
with no cycles.  In a slight abuse of notation, we shall use 
the symbol $\mathbb{T}$ for both the tree graph and the set 
of its vertices.   The root is a particular vertex which we 
denote $0 \in \mathbb{T}$.  For each $x\in \T$ we denote by 
$|x|$ the number of edges in the unique path connecting it 
to the root.   In a regular tree, as those considered here, each 
vertex other than the root has $K+1$ neighbors, one towards 
the root and $K$ in what we refer to as the forward direction. 
The set of the forward neighbors of $x$ is denoted by
$\mathcal{N}_x^+$.   
We say that $ y \in \T $ is in the future of $ x \in \T $   
if the path connecting $ y $ and the root runs through $ x $.  
The subtree consisting of all the vertices in the
future of $ x $, with  $x$ regarded 
as its root,  is denoted by $\mathbb{T}_x $.\\

The symmetries referred to in  {\bf A1} 
are associated with 
endomorphisms of the rooted tree. 
These are mappings  $ s:  \T\to \T $ 
preserving the adjacency  relation and the 
orientation away from the root, 
i.e., neighboring vertices are mapped onto neighboring 
vertices, and
if  $ x $ is in the future of $ y $ then $ sx $ is in the future of $ sy $.
To each such endomorphism corresponds a transformation  
$S: \C^{\T} 
\to \C^{\T}$ defined by $(S\omega)_x := \omega_{s x}$ for all $ x \in \T $. 
A probability 
measure $ \nu $ on $ \C^\T $  is said to be {\em stationary} if for all such mappings and all bounded measurable 
$ 
F: \C^\T \to \C $
\begin{equation}
        \int_{\C^\T } F(S \omega) \, \nu(d \omega) =  
\int_{\C^\T } 
F( \omega ) \, \nu(d \omega) \, .
\end{equation}

The weak correlation condition required in {\bf A3}
is the subject of the following
\begin{definition}
A probability measure $ \nu $ on $ \C^\T $ is said to be {\em weakly
correlated} if
there exists some $\kappa \in(0, 1]$ such that
for any pair of vertices $x \neq y $, which are common forward neighbors of some vertex, and any pair of bounded
measurable functions $F, G: \, \C^\T \to [0,\infty)$, one of which is
determined by the values over the forward subtree
$ \T_x $ and the other determined by the values of over the forward subtree $\T_y$,
\begin{equation}\label{ass:wc}
\int_{ \C^\T } F(\omega)\, G(\omega) \, \nu(d\omega) \geq \, \kappa 
\; \int_{ \C^\T
} F(\omega) \, \nu(d\omega) \; \int_{ \C^\T } G(\omega) \, \nu(d\omega).
\end{equation}
\end{definition}

By standard approximation arguments it suffices to test
\eqref{ass:wc} for bounded continuous $ F $, $ G $.\\

Clearly, the collection of
probability measures $ \mathbb{P} $ on $ \R^\T $
satisfying {\bf A1} and {\bf A3}
includes the case where $\{ \omega_{x}\}_{x\in \T}$ form iid random 
variables. 
Let us also note that  if $ \P $ is
stationary and weakly correlated then $ \P $ is ergodic. As a consequence, the ac spectrum of $ H_\lambda(\omega) $ coincides with a 
deterministic set, cf.\ \cite{CaLa90,PF,Klein95}.

\subsection{Relation with previous results} 

The topic of Anderson 
(de)localization on tree graphs goes
back to Abou-Chacra, Anderson 
and Thouless \cite{Abou73,Abou74}.  They noted that a 
``self-consistent'' approach, which for general graphs can be viewed 
as an approximation, is exact for Cayley trees and used it to explore the location of the mobility edge for small disorder.  
The subject was further 
studied in \cite{MiFy91,MilDer93}.
In particular, Miller 
and Derrida \cite{MilDer93}
argued that  for energies within  the 
spectrum of the unperturbed operator delocalized eigenfunctions 
should persist under weak 
disorder.  For a summary of these findings   
see~\cite{Klein95}.\\

There have also been rigorous results on this topic.   
Localization, in the  sense of existence of pure point spectrum, 
was proven at extreme energies, and at all energies 
for large 
values of the disorder parameter $\lambda$, see~\cite{AM}.  
It was also shown in \cite{A2} that for small $\lambda$ and $ U = 0 $ the operator 
$ H_\lambda(\omega) $ has only pure point spectrum for energies $|E| > K+1$.  
It should be noted that this range does not include the full resolvent set of the unperturbed operator, since 
\begin{equation}\label{eq:Tfree}
          \sigma(T) = \sigma_{\rm ac}(T) = \big[-2 
\sqrt{K},2 \sqrt{K} \big].
\end{equation} 
Concerning delocalization, 
which is the subject of this note, Klein \cite{Klein95,Klein98} 
established the stability of the ac spectrum
in the case $ U = 0 $ and $\{ \omega_x\}_{x\in \mathbb{T}} $ iid random variables.
Using supersymmetric representations he showed
  that for any  $ 0 < 
|E| < 2 \sqrt{K} $ there exists
  $\lambda(E) > 0 $ such 
that
\begin{equation} \label{eq:Klein}
\sup_{|\lambda|< \lambda(E)}\, 
\sup_{|E'|< E } \, \sup_{\eta > 0 }
\;\mathbb{E}\left[\big|
\left\langle \delta_0, (H_{\lambda}(\cdot)-E'-i\eta)^{-1} \, 
\delta_0\right\rangle
  \big|^2\right] <
\infty .
\end{equation}
In particular, (\ref{eq:Klein}) implies  that for small $\lambda$ the almost sure
spectrum is 
\emph{purely} ac in an energy range which 
is contained in $ \sigma_{\rm ac}(T) $ as shown in \cite{Klein98}. Moreover,
the states in this 
energy range exhibit super-ballistic-transport 
behavior  \cite{Klein96}. \\

The results presented here address issues similar 
to those discussed in~\cite{Klein98}.  We do not pursue the question  whether 
the ac spectrum is pure in the intervals under study.   
However, the approach we present is  quite different from 
the technique used in the above mentioned works, 
and the result applies to more general situations.

\section{A criterion for the stability of the ac spectrum}\label{sec:radstat}

The argument which proves our main result, Theorem~\ref{thm:main}, 
yields a continuity criterion of a somewhat greater generality. 
To present it, we shall  frame the discussion in the context of 
radially stationary potentials.   For this purpose, we let 
$ (\Xi,p) $ be a probability space on which there is measure preserving 
ergodic mapping $ \mathcal{S}: \Xi \to \Xi $.  
Every measurable function $ u :  \Xi \to \R $ generates through
\begin{equation}\label{eq:radstat}
        U_x(\theta) = u(\mathcal{S}^{|x|}\theta), \qquad \mbox{for all}\quad  x \in \T, \; \theta \in \Xi,
\end{equation} 
a potential on the tree, which is radial and stationary under radial shifts. We will subsequently refer to potentials $ U(\theta) $ of the 
form \eqref{eq:radstat} as \emph{radially stationary} and assume tacitly that the function $ u $ is log-integrable: 
\begin{equation}
         \int_\Xi \log( 1+ | u(\theta)|) \, p(d\theta) < \infty \, .
\end{equation}

To include the  radially $ \tau$-periodic  potentials in the 
above setup, one may take   
$ \Xi = \{1, \dots, \tau\} $, with  $ p $ the equidistribution 
among the $ \tau $ integers, and  $ \mathcal{S} $ the shift  
$ \mathcal{S} \theta := \big(\theta + 1\big) \mod \tau $.  \\

\subsection{The stability criterion}
Any radially stationary potential gives rise to a \emph{radially stationary Schr\"odinger operator}
\begin{equation}
        H_0(\theta) := T + U(\theta) \qquad \mbox{on $ \ell^2(\T) $.}
\end{equation}
Some basic facts on the spectral properties of such operators are collected in Appendix~\ref{app:proof}. 
As a generalization of \eqref{eq:H} we consider weak perturbations of such operators
by a random potential, i.e.,
\begin{equation}
         H_\lambda(\theta,\omega) :=    H_0(\theta) + \lambda V(\omega) \qquad \mbox{on $ \ell^2(\T)$.} 
\end{equation}
For the statement of our stability criterion it is important to note that there exists a Borel set
$ \Sigma_{\rm ac}(0) \subseteq \R $ such that $
        \sigma_{\rm ac}\big(H_0(\theta)\big) = \Sigma_{\rm ac}(0) $ for almost all $ \theta \in \Xi $, cf.\ Proposition~\ref{prop:1d2}.
Moreover, due to ergodicity there exists a Borel set $ \Sigma_{\rm ac}(\lambda) \subseteq \R $ such that
\begin{equation}
                \sigma_{\rm ac}\big(H_\lambda(\theta,\omega)\big) = \Sigma_{\rm ac}(\lambda) 
\end{equation}
for almost all $ ( \theta,\omega) $.\\

In the proof of Theorem~\ref{thm:main} a significant role will be played by 
considerations of functions $\Gamma  $ with $ \Im \Gamma \geq 0 $ satisfying 
the following co-cycle condition 
\begin{equation}\label{eq:cocycle}
 \Gamma(\theta) = \frac{1}{U_0(\theta) - E - K \, \Gamma(\mathcal{S}\theta)} \, . 
        \end{equation}
It provides an alternative formulation of the Schr\"odinger equation 
for a covariant eigenfunction.  
As will be discussed below, the co-cycle~\eqref{eq:cocycle} has at least one solution with $ \Im \Gamma \geq 0 $ 
in the energy range $ \Sigma_{\rm ac}(0) $.
Its uniqueness in case of radially periodic $ U $ is proven in Proposition~\ref{persol} of Appendix~\ref{app:proof}. 

\begin{definition} 
The Schr\"odinger co-cycle~\eqref{eq:cocycle}  is said to 
admit a unique solution in an energy range $I\subset \R$ 
 if for Lebesgue-almost all $ E \in I $ 
there exists a  unique measurable function 
$ \Gamma: \Xi \to \C $ 
satisfying \eqref{eq:cocycle} and  $ \Im \Gamma(\theta) \geq 0 $ for almost every $ \theta \in \Xi $. 
\end{definition}

As is apparent from the argument, our proof of 
Theorem~\ref{thm:main}, establishes the following  somewhat 
more general statement.  

\begin{theorem}\label{thm:main2}
Let $ U(\theta) $ be radially stationary for which the spectrum of 
$ H_0(\theta) $ has an absolutely continuous component, and 
let $ V(\omega) $ be a random potential satisfying {\bf A1}, {\bf A2}, 
and {\bf A3}.   
Then  a sufficient condition for the continuity of the ac spectrum of   
 $ H_\lambda(\theta,\omega) $    
 in the sense expressed in Theorem~\ref{thm:main},  is that  the Schr\"odinger co-cycle~\eqref{eq:cocycle} 
admits a unique solution in the energy range $\Sigma_{\rm ac}(0) $.
\end{theorem}

It should be noted that if $ U(\theta) $ is non-deterministic then $ \Sigma_{\rm ac}(0) = \emptyset $ 
due to the relation of $ \Sigma_{\rm ac}(0) $ to the ac spectrum of a one-dimensional operator 
discussed in Appendix~\ref{app:proof} 
and
Kotani theory \cite{Kot83,Sim83}. 
Examples of deterministic
potentials are periodic or almost-periodic ones. 
For a further comment on the latter case, see Section~\ref{sec:disc}.

\subsection{Outline of the proof}

The analysis on trees is often more accessible than on other graphs
since various quantities computed at the tree root satisfy recursion
relations.  These   relate the quantity at the root to the
corresponding  counterparts on the subtrees to which the tree breaks
upon the removal of the root site.
We make use of such a relation for the  diagonal elements of the
forward resolvents:
\begin{equation}\label{eq:defgam}
\Gamma_x( \lambda, z,\theta,\omega ) := \left\langle \delta_x \, ,
\big(H_{\lambda}^{\T_x}(\theta,\omega) -z\big)^{-1} \delta_x \right\rangle,
\end{equation}
where
$H_{\lambda}^{\T_x}(\theta,\omega)$ is the restriction
of the operator
$H_{\lambda}(\theta,\omega) $ to the Hilbert space $ \ell^2( \T_x) $ over
the forward tree graph
$ \T_x \subset \T $.
The above is well defined for any $z \in \mathbb{C}^+ := \{ z \in \mathbb{C}: \Im z >0 \} $,
and through the Herglotz property the limit
\begin{equation}
    \Gamma_x( \lambda, E+ i0 ,\theta, \omega ) := \lim_{\eta\downarrow 0}
\Gamma_x( \lambda, E+ i \eta ,\theta, \omega )
\end{equation}
exists for  Lebesgue-almost every $E\in\R$, cf.\ Appendix~\ref{app:bounds}.\\

The  forward resolvents play a
diagnostic role for the problem considered here.  In particular, the
density of the ac component of the spectral measure of $
H_\lambda(\theta,\omega) $ associated with $\delta_0 \in \ell^2(\T) $ is given by $ \pi^{-1} \Im
\Gamma_0( \lambda, E+ i0 ,\theta,\omega ) $.

It is a significant observation that the above quantities also play
another role.
The products yield the off-diagonal Green function \cite[Eq.~(2.8)]{Klein98},
\begin{equation}\label{eq:off-diag}
          \left\langle \delta_0, (H_{\lambda}(\theta,\omega)-z)^{-1}\,
\delta_x \right\rangle = \prod_{j=0}^{|x|}\, \Gamma_{x_j}(\lambda, z,\theta,\omega),
\end{equation}
where  $ x_j $, with $ j=0, \dots, |x| $, denote the vertices along
the unique path joining the root $  0 \, (=: x_0) $
and $ x \, (=: x_{|x|} ) $.

Fundamental to our discussion is the recursion  relation which the
forward resolvents
are well know to satisfy \cite{Abou73,MilDer93,Klein98,FHS_04}.  For each
   $\lambda \in \R$, $ z \in \mathbb{C}^+ $,
$ \omega \in \R^\T$ and at each vertex $ x \in \mathbb{T} $, one has
\begin{equation} \label{recur}
     \Gamma_x(\lambda,z,\theta,\omega)=  \Big( U_x(\theta) + \lambda V_x(\omega) - z - \sum_{y \in
    \mathcal{N}_x^{+}} \Gamma_y(\lambda,z,\theta,\omega)\, \Big)^{-1},
\end{equation}
where $\mathcal{N}_x^{+}$ is the set of the forward neighbors of $x$. For $ \lambda = 0 $ and $ E \in \R $ 
this relation boils down to \eqref{eq:cocycle} 
due to covariance property
\begin{equation}
        \Gamma_x(0,z,\theta) = \Gamma_0(0,z,\mathcal{S}^{|x|}\theta) \, .
\end{equation}
For almost every $ E \in \Sigma_{\rm ac}(0) $ one measurable solution of \eqref{eq:cocycle} with values in $ \C^+ $ is thus provided by 
$ \Gamma_0(0,E+i0,\theta) $, the forward resolvent corresponding to $ H_0(\theta) $. The issue in the additional assumption of 
Theorem~\ref{thm:main2} is therefore the uniqueness of this solution.\\

The main part of our analysis is to show that 
the forward resolvents
converge in a certain distributional sense to their unperturbed counterparts.  
To do so, we first prove that their
 distribution is sharp in the sense that 
\begin{equation} \label{eq:fluct}
\Gamma_x(\lambda,E+i \eta,\theta,\omega) \ = \
\Phi_x(E,\theta) \; \big[ 1+ o_x(E,\theta,\omega;\lambda,\eta)\big]
\end{equation}
with some $  \Phi_x(E,\theta) $ 
 which does not depend on $\omega$ and certain
$o_x(E,\theta, \omega;\lambda,\eta)$ which vanish in the distributional
sense for $\lambda,\eta \to 0$. \\

A key step in the derivation of \eqref{eq:fluct} is the proof of the
corresponding statement for just the imaginary part, $ \Im \Gamma_x $,
where we suppress the dependence on $ (\lambda,E+i\eta,\theta,\omega) $.
The starting point for this is the relation:
\begin{equation}\label{imrecur}
        \log \Big( \Im \Gamma_x\Big)  \ = \  \log \Big(  K   \,
\big|\Gamma_{x}\big|^2  \Big)
       + \log\Bigg(\frac{\eta}{K} +
       \frac{1}{K} \sum_{y\in \mathcal{N}_{x}^+} \Im \Gamma_y\Bigg),
\end{equation}
which follows from \eqref{recur}.  The mean value of the first term
on the right may be regarded as a Lyapunov exponent on the tree
which  vanishes at $\lambda,\eta=0$  for most $E\in\Sigma_{\rm
ac}(0)$. A relevant  observation here is that while  this term may
exhibit a rather erratic dependence on $ \lambda $ and $E$ for $\eta
 = 0$,
its integrals over $E$ form continuous functions
of  $\lambda,\eta$.  

The second term calls for an application of
the Jensen inequality:
\begin{equation}\label{Jensen}
\log\Bigg(\frac{1}{K} \sum_{y\in \mathcal{N}_{x}^+} \Im \Gamma_y
\Bigg) \ \ge \  \frac{1}{K} \sum_{y\in \mathcal{N}_{x}^+} \log \Im
\Gamma_y  \,,
\end{equation}
where the inequality is strict unless $\Im \Gamma_y  $ coincide for all  $y\in
\mathcal{N}_{x}^+ $.
Under the weak-correlation assumption {\bf  A3}, the above
considerations lead to the conclusion that the distributions of $\Im
\Gamma_x $ are of vanishing relative width. This is quantified below
through a more thorough discussion of that notion  -- 
on which we expand in Appendix~\ref{App:error} --  
and
a strengthened version of the Jensen inequality.\\

Putting the above arguments together we conclude the analog of \eqref{eq:fluct} for
$\Im \Gamma_x$ and use this to deduce the sharpness of the
distribution of $\Gamma_x$ itself.
Finally, upon substituting \eqref{eq:fluct} into the recursion
relation \eqref{recur}, we conclude that $ \Phi_x(E,\theta)  $ satisfies the
same equation as $\Gamma_x(0,E+i0,\theta)$.  Since in its dependence on $x$, 
$\Phi_x(E,\theta)$ is radial and covariant, i.e., $
        \Phi_x(E,\theta) = \Phi_0(E,\mathcal{S}^{|x|}\theta) $,
our main result follows from the uniqueness of solutions of \eqref{eq:cocycle}.


\section{A Lyapunov exponent and its continuity}

We shall refer to the following quantity as the Lyapunov exponent for
the Schr\"o\-dinger operator \eqref{eq:H} on the tree
\begin{equation}\label{def:Lya}
          \gamma_\lambda(z) :=
         -  \int_\Xi
        \mathbb{E}\Big[\log\Big( \sqrt{K} \; \big|
\Gamma_{0}(\lambda, z,\theta,\cdot)  \big|\Big)\Big]\, p(d\theta) \, .
\end{equation}
Below, we list some basic facts:
\begin{enumerate}
\item The Lyapunov exponent $ \gamma_\lambda(z) $ is a harmonic
    function of $ z \in \mathbb{C}^+ $ as it is the negative real part of the
Herglotz function 
          \begin{equation}\label{def:w}
          w_\lambda(z):= \int_\Xi \mathbb{E}\Big[\log\Big( \sqrt{K} \; \Gamma_{0}(\lambda, z,\theta,\cdot) \Big)\Big]\, p(d\theta) \, .
          \end{equation}
          Assumption~{\bf A2} and Lemma~\ref{lem:Gup} ensure that $
w_\lambda(z) $ and $ \gamma_\lambda(z) $ are well defined.
   Moreover, by the symmetry assumption~{\bf A1}, $w_{\lambda}(z)$ remains unchanged if $ \Gamma_0 $ is replaced by $ \Gamma_x $.
\item In view of
the relation \eqref{eq:off-diag},
$ \gamma_\lambda(z)$  describes the typical decay rate of the Green's
function  along a ray, normalized so that in the absence of
significant fluctuations between different branches
   $\gamma_\lambda(z)>0$ assures that the Green function is square 
integrable. The reader is cautioned, however, that
          in contrast to one dimension, for random potentials on the 
tree the ac spectrum
does not coincide with the essential closure of the set of energies
          on which this Lyapunov exponent vanishes. This issue is
further discussed in Appendix~\ref{App:thouless}.
\end{enumerate}

Some of the properties of the Lyapunov exponent which are
of immediate relevance for our discussion 
are summarized in the following statement.

\begin{theorem}\label{thm:cont}
    Let $U(\theta)$ be radially stationary, and $V(\omega)$
satisfy {\bf A1} and {\bf A2}. Then:
\begin{enumerate}\itemsep0.5ex
\item The Lyapunov exponent is positive  
          $\gamma_\lambda(z) > 0$ for all $z \in \C^+$, and satisfies 
        \begin{equation}\label{eq:asym}
         \lim_{\eta \to \infty} \frac{\gamma_\lambda(E+i\eta)}{\eta} = 0 \, .
        \end{equation} 
\item  For $ \lambda = 0 $ the Lyapunov exponent vanishes on the ac spectrum of $H_0(\theta)=T+U(\theta)$:  
        \begin{equation}
        \gamma_0(E+i0) = 0 
\end{equation}
for Lebesgue-almost every $E\in \Sigma_{\rm ac}(0)$.
\item
    For any bounded Borel set 
    $ I \subset \R$ the integral
    $\int_I \gamma_\lambda(E+i\eta) \, dE$ is continuous in $
(\lambda,\eta)\in \mathbb{R}\times [0,\infty)$.
\item For any bounded Borel set 
    $  I \subseteq \Sigma_{\rm ac}(0) $ one has
    \begin{equation}\label{eq:cont}
      \lim_{\substack{\lambda \to 0 \\ \eta \downarrow 0}}\; \int_I
\gamma_\lambda(E+i\eta) \, d E  =  0 .
    \end{equation}
\end{enumerate}
\end{theorem} 
\begin{proof}
1. The positivity of $ \gamma_\lambda $ on $ \C^+ $ can be seen through \eqref{imrecur} and \eqref{Jensen}. The statement about 
the asymptotics derives from (\ref{Gup}) in Appendix~\ref{app:bounds}.\\[0.5ex]
2. The vanishing of $\gamma_0 $ on $ \Sigma_{\rm ac}(0)$ forms part of Proposition~\ref{prop:1d2} in 
Appendix~\ref{app:proof}. \\[0.5ex]
3. As a positive harmonic function which satisfies the 
asymptotics~\eqref{eq:asym}, 
$\gamma_{\lambda}(\cdot +i \eta) $ with $ (\lambda, \eta ) \in \R \times [0,\infty) $ can be represented as
\begin{equation}
         \gamma_{\lambda}(z +i \eta)  = \int_\R \frac{\Im z}{|E -z |^2} \, \sigma_{(\lambda,\eta)}(dE) \, ,
\end{equation}
where the Borel measure $ \sigma_{(\lambda,\eta)} $ is unique and satisfies
$ \int_\R (E^2+1)^{-1} \sigma_{(\lambda,\eta)}(dE) < \infty $, see~\cite{Duren}. 
Thanks to the Hergoltz property of $w_\lambda $, the harmonic conjugate of 
$ \gamma_{\lambda}(\cdot +i \eta) = - \Re w_\lambda( \cdot + i \eta ) $ has a 
definite sign and hence locally integrable boundary values \cite[Thm.~1.1]{Duren}. This implies that $ \sigma_{(\lambda,\eta)} $ is purely ac
\cite[Thm.~3.1 \& Corollary~1]{Duren}, and for all $ (\lambda, \eta ) \in \R \times [0,\infty) $, one has 
\begin{equation}
        \sigma_{(\lambda,\eta)}(I) = \int_I \gamma_\lambda(E + i \eta) \, dE  \, .              
\end{equation}
The asserted continuity thus follows from the vague continuity of the measure $ \sigma_{(\lambda,\eta)} $ in 
$ (\lambda, \eta ) \in \R \times [0,\infty) $, see~\cite{MBauer}. By Proposition~\ref{lemma:critcont} below, 
a sufficient condition for the latter is the (pointwise) continuity of 
$ \gamma_{\lambda}(z +i \eta) $ for all $ z \in \C^+ $. 
This pointwise convergence follows from the (weak)  resolvent convergence
\begin{equation}\label{eq:srconv}
    \lim_{\substack{\lambda' \to \lambda \\ \eta' \to \eta }}
   \Gamma_x(\lambda', z + i \eta',\theta,\omega) =
   \Gamma_x(\lambda, z + i \eta,\theta, \omega) 
\end{equation}
for all $ \Im z > 0 $ and all $ x \in \T $, $ \lambda \in \R $, $ \eta \in [0,\infty) $, $ \theta \in \Xi $, $ \omega \in \R^\T $, 
together with the 
dominated convergence theorem, which is applicable thanks to
(\ref{Gup}) of Appendix~\ref{app:bounds}.\\[0.5ex]
4. This is an immediate consequence of 2.\ and 3. \qed
\end{proof}
The previous proof relied on the following convergence statement, which we recall from \cite[Prop.~4.1]{HLMW01}. 

\begin{proposition}\label{lemma:critcont}
Let $ \sigma , \sigma_n $ be non-negative Borel measures satisfying
$ \int (E^2+1)^{-1} \sigma(dE) < \infty $ and similarly for 
$ \sigma_n  $. Assume that for all $ z \in \C^+ $
\begin{equation} 
        \lim_{n \to \infty}  \int \frac{1 }{|E - z |^{2}} \, \sigma_n(dE) =  \int \frac{1}{|E - z |^{2}}\, \sigma(dE) \, ,
\end{equation}
then $ \lim_{n \to \infty} \sigma_n = \sigma $ in the sense of vague convergence.
\end{proposition}

For a definition of vague convergence and its implications,
see \cite{MBauer}.


\section{Fluctuation bounds}
An important tool for our analysis is the following strengthened version of the
Jensen inequality for the logarithm.
\begin{lemma}\label{ref:boost}
          Let $ K \geq 2 $. Then for any collection $ (X_j)_{j=1}^K $
of positive numbers
          \begin{equation}
                   \log \left( \frac{1}{K} \sum_{j=1}^K X_j \right)
\geq \frac{1}{K} \sum_{j=1}^K \log X_j
                  + \frac{1}{2K(K-1)} \sum_{i \neq j}  \left(\frac{X_i
- X_j}{X_i + X_j}\right)^2.
          \end{equation}
\end{lemma}
\begin{proof}
          We expand the average inside the logarithm into an average
over pairs and use Jensen's inequality to obtain
          \begin{equation}
                    \log \left( \frac{1}{K} \sum_{j=1}^K X_j \right)
                  \geq \frac{1}{K(K-1)} \sum_{i \neq j} \log \left(
\frac{X_i + X_j}{2} \right).
          \end{equation}
          This reduces the claimed inequality to the case $ K = 2 $.
For a proof of the latter let
          $ \xi_j := 2 X_j / (X_1 + X_2) $, which takes values in $
[0,2] $, and let
          $ f(\xi) := - \log \xi + ( \xi - 1 ) $.
          Using the  symmetry $ 2 - \xi_{1/2} = \xi_{2/1} $, we have
          \begin{equation}
                  \log\left(\frac{X_1 + X_2}{2}\right) - \frac{1}{2}
\left( \log X_1 + \log X_2 \right)
                   =  \frac{1}{4} \sum_{j=1}^2 \big( f(\xi_j) + 
f(2-\xi_j) \big).
          \end{equation}
By elementary arguments
          $ f(\xi) + f(2-\xi) \geq  (\xi-1)^2 $ for all $ \xi \in (0,2]
$, which finally implies the assertion of the lemma.
\qed\end{proof}
The above improvement of the Jensen inequality for $\log X$ is
substantial unless the empirical distribution of $\{ X_j \}$ is
narrow in a sense which is quantified as follows.

\begin{definition}\label{def:relwitdth}
          For $ \alpha \in (0,1/2] $  the 
 \emph{relative $ \alpha $-width} of a probability measure $\nu$ 
 on $(0,\infty)$  is given by
          \begin{equation}
                  \delta(\nu,\alpha) := 1 - \frac{\xi_-(\nu,
\alpha)}{\xi_+(\nu, \alpha)}.
          \end{equation}
where 
 \begin{eqnarray} \label{eq:defxi}
\xi_-(\nu, \alpha) &=& \sup \{\, \xi \, :\,  \nu[0,\xi) \le \alpha\} \nonumber 
\\
 \xi_+ (\nu, \alpha) &=&  \inf \{\,  \xi \, :\,  \,  \nu(\xi,\infty) \le \alpha\} \, .
\end{eqnarray}
For a random variable $X$ we denote 
\begin{equation}
 \delta(X,\alpha) \  := \   \delta(\nu_X,\alpha) 
\end{equation}
where $\nu_X$ is the probability distribution of $X$, which is 
defined by: $\nu_X(A) =\P(X\in A)$.  
\end{definition}

Thus, small $  \delta(X,\alpha) $ means that the distribution of $ X $
is sharp in the sense expressed in \eqref{eq:fluct}.
Some useful observations about the composition laws for  the relative
widths of sums and products of random variables are presented in
Appendix~\ref{App:error}.\\

We shall now apply the above tools to show that the vanishing of
the Lyapunov exponent implies the sharpness of the distributions of
both $ \Im \Gamma_x $ and $ | \Gamma_x | $.
\begin{theorem}\label{thm:imfoc} Let $ U(\theta) $ be radially stationary, and $ V(\omega) $
satisfy {\bf A1}, {\bf A2} and {\bf A3}. Then for any $ x \in \T $, $ \lambda \in
\R $, $ z \in \C^+ $ and $ \alpha \in (0,1/2] $:
    \begin{align}\label{eq:imfoc1}
             \int_\Xi \, \delta\big(\Im
\Gamma_{x}(\lambda,z,\theta,\cdot),\alpha \big)^2 \, p(d\theta) & \leq
          \frac{8}{\kappa \, \alpha^2} \, \gamma_\lambda(z), \\[1.5ex]
             \left[ \int_\Xi \, \delta\big( \big|
\Gamma_{x}(\lambda,z,\theta,\cdot) \big|^2 ,\alpha \big) \, p(d\theta) \right]^2
          & \leq \frac{32\,  (K+1)^2  }{\kappa \, \alpha^2}
\,\gamma_\lambda(z). \label{eq:imfoc2}
    \end{align}
Here $ \kappa $ is the constant appearing in the weak-correlation
condition \eqref{ass:wc}.
\end{theorem}
\begin{proof}
For a proof of \eqref{eq:imfoc1}, we start from (\ref{imrecur}) which
implies the inequality
\begin{equation}\label{eq:averimrec}
        \mathbb{E}\Big[\log \big( \Im \Gamma_x \big) \Big] - 2 \,
\mathbb{E}\left[\log \left( \sqrt{ K } \,
|\Gamma_{x}|  \right) \right]
       \geq \mathbb{E}\left[\log\Bigg(\frac{1}{K} \sum_{y\in
\mathcal{N}_{x}^+} \Im \Gamma_y \Bigg)\right],
     \end{equation}
where again we will suppress the dependence on $ (\lambda,z,\theta,\omega) $.
Lemma~\ref{ref:boost} yields a lower bound for the right side of
(\ref{eq:averimrec}) consisting of
a sum of two terms, $S_1$ and $S_2$. The first is
    \begin{equation} \label{eq:aoa}
        S_1 \ :=\  \mathbb{E}\Bigg[ \frac{1}{K}
          \sum_{y\in \mathcal{N}_{x}^+} \log \big( \Im \Gamma_y \big) 
\Bigg] \, ,
    \end{equation}
which, upon averaging over $ \theta $ and using {\bf A1},
will cancel the first term on the right side of \eqref{eq:averimrec}.
    The second term is
\begin{equation}
      S_2\ :=\   \frac{1}{2K (K-1)} \sum_{y, y'\in \mathcal{N}_{x}^+}
\; \mathbb{E}\left[  \left(\frac{\Im \Gamma_y
          -\Im \Gamma_{y'}}{\Im \Gamma_y+ \Im \Gamma_{y'} }\right)^2   \right].
    \end{equation}
The variables $\Gamma_y$ and $\Gamma_{y'}$ appearing in the above
summation $ y, y' \in  \mathcal{N}_x^{+}$
are identically distributed due to the symmetry  implied by the 
stationarity condition {\bf A1}.  
Moreover, by the weak-correlation condition {\bf A2} the measure
describing their joint distribution is bounded below by $\kappa
\times$ the product measure which describes two independent
copies sampled from the common distribution.
Using Lemma~\ref{lemma:wcorr} of Appendix~\ref{App:error}, we find
\begin{equation}\label{eq:2terms}
S_2 \    \geq    \frac{\kappa \alpha^2}{4}
           \left[ \delta\big(\Im \Gamma_{y},\alpha \big) \right]^2
    \end{equation}
Combining the terms $S_1$ and $S_2$, and averaging
\eqref{eq:averimrec} over $ \theta $,
one  arrives at (\ref{eq:imfoc1}).

    The second assertion \eqref{eq:imfoc2} follows from squaring the
following inequality,
    \begin{equation} \label{db}
           \int_\Xi \delta \left( \big| \Gamma_{x}(\lambda, z,\theta,\cdot
)\big|^2, \alpha \right) \, p(d\theta) 
           \leq  2 \, \int_\Xi
          \delta \Big( \Im \Gamma_{x}(\lambda, z ,\theta,\cdot), \frac{
\alpha}{K+1} \Big) \, p(d\theta) \, ,
          \end{equation}
          applying the
          Jensen inequality, and inserting (\ref{eq:imfoc1}). For a
proof of (\ref{db}), we employ the recursion relation (\ref{recur})
and
          Lemma~\ref{lemma:delta}, which for any $p+q=1$ yields
          \begin{align}
          \delta \left( | \Gamma_x|^2, \alpha \right)
          & \leq \delta \left(
          \Im \Gamma_x, p \alpha \right) +
          \delta \Bigg( \sum_{y\in \mathcal{N}_x^+}  \Im \Gamma_y, q
\alpha \Bigg)  \notag \\
          & \leq   \delta \left( \Im \Gamma_x, p \alpha \right) + 
\delta \left(   \Im \Gamma_y, \frac{q \alpha}{K} \right).
          \end{align}
          In the last inequality, we have used the fact that
          $\Gamma_y$ is identical in distribution to $\Gamma_{y'} $ for
any $ y$, $ y' \in \mathcal{N}^+_x $.
          Setting $p = (K+1)^{-1}$ and averaging over $\theta $ completes the proof of (\ref{db}).
\qed\end{proof}

\section{Distributional convergence of the forward resolvents}

Our goal now is to establish that in a certain distributional
sense, on which
more is said below,
\begin{equation}
\label{eq:limit}
\Gamma_x(\lambda, E+i \eta, \theta,\omega)  \ \mathop {\longrightarrow}_{\lambda, \eta \to 0 }^{\mathcal D}  \ \Gamma_x(0, E+i0,\theta)
\end{equation}
where the quantity on the right side is a forward resolvent of $H_0(\theta)$.  The underlying reasoning
for \eqref{eq:limit} is the observation that the fluctuation bounds of
Theorem~\ref{thm:imfoc} and the information on the
Lyapunov
exponent in Theorem~\ref{thm:cont} imply that the prelimit in \eqref{eq:limit}
exhibits very weak
dependence on
$\omega$ for small $\lambda $ and $ \eta$.  At the same time, those
quantities satisfy
a recursion relation which is close to \eqref{eq:cocycle}.  By assumption this equation has a unique solution taking values 
in $ \C^+ $.

There are a number of gaps in the above narrative which
need to be addressed in the proof:
\begin{enumerate}
\item Concerning
the Lyapunov exponent $\gamma_\lambda(E+i\eta)$,
it is only known that the integral over $E$ tends to zero
in the joint limit  $\lambda, \eta \to 0 $ -- not
that it tends to zero on some set of
energies $E $.
\item The fluctuation bounds imply the narrowing of
the distribution of $|\Gamma_x|$ and $ \Im\Gamma_x$, however, the limiting value of $\Gamma_x\in \C^+$ could range over 
two distinct points.
\item The
narrowing in the distribution refers only to the dependence of
$\Gamma_x(\lambda, E+i \eta, \theta, \omega)$ on $\omega$ at fixed
$\lambda,\eta $.   That still leaves room for some rather erratic
dependence on the parameters $\lambda,\eta$.   
\item  The uniqueness of the solution holds only for the limiting equation and
within the class of perfectly radial and covariant functions of $|x|$. However, we deal with quantities which may 
still include additional
randomness and for which the symmetry holds only in the distributional sense.
\end{enumerate}

In order to bypass the
first-mentioned limitation, the convergence
\eqref{eq:limit}
is derived below in the distributional sense with respect
to the joint dependence on $(E,\omega)$.  The latter are
distributed by the product measure
$\mathcal D := \mathcal{L}_I\otimes  \mathbb{P} $
on $I\times \R^\T$, where $\mathcal{L}_I$
is the Lebesgue measure
restricted to $I \subseteq \Sigma_{\rm ac}(0) $.
In order to
address the joint values of the entire collection of variables, we
regard
$(E, \Gamma)$ as taking values in the
product space
$ I \times \mathbb{C}^\T  $
and consider the family
of  finite measures
$ \mu_{(\lambda, \eta)}^{(\theta)} $ induced on
this space by the image  of
$\mathcal{L}_I \otimes \mathbb{P} $ under the
mapping
\begin{equation}\label{def:mun}
         (E,\omega) \mapsto \left(E,\big\{
\Gamma_x(\lambda,E+i\eta,\theta,\omega) \big\}_{x \in \T} \right) \, ,
\end{equation}
where $ \theta\in \Xi $ and $ (\lambda,\eta)$ are indexing parameters with
values in $\mathbb{R} \times (0,\infty) $ or $ \{(0,0)\} $.
These mappings and the corresponding measures
are well-defined even along the boundary $ \eta = 0 $.\\

In the above terminology, we will establish the following
\begin{theorem}\label{thm:meas}
Let $ U(\theta)$ be radially stationary, assume that the co-cycle~\eqref{eq:cocycle} admits a unique solution
on $\Sigma_{ac}(0) $, and let $ V(\omega) $ satisfy {\bf A1},
{\bf A2} and {\bf A3}. Moreover, let $  I \subseteq \Sigma_{ac}(0) $ be a bounded Borel set. 
Then the measures on $ I \times \mathbb{C}^\T  $ which describe the joint
distribution of $(E,\Gamma)$ induced from   $
\mathcal{L}_I\otimes \mathbb{P}$ by the mapping \eqref{def:mun},
satisfy for almost all $ \theta \in \Xi $:
        \begin{equation}\label{eq:weakconv}
        \lim_{\substack{\lambda\to 0 \\ \eta \downarrow 0}} \;
\mu_{(\lambda,\eta)}^{(\theta)}= \mu_{(0,0)}^{(\theta)}
        \end{equation}
        in the sense of weak convergence.

\end{theorem}

Let us note a number of elementary properties of the measures
$ \mu_{(\lambda, \eta)}^{(\theta)}$ and some related observations:
\begin{enumerate}
\item   The limiting measure $ \mu_{(0,0)}^{(\theta)}$ is concentrated on the
graph of the function
        $  E \mapsto \Gamma(0,E+i0,\theta) \in \C^\T $. Accordingly, it has the product form 
        \begin{equation}
                \mu_{(0,0)}^{(\theta)}(dE \,d\Gamma) = dE \,  \otimes  \delta_{\Gamma(0,E+i0,\theta)}(d\Gamma).
        \end{equation}
\item For almost every $ E\in I $ the conditional measure $ \mu_{(\lambda, \eta)}^{(E,\theta) }$
equals the image of the probability measure $ \mathbb{P} $ under the
mapping $ \omega \mapsto \Gamma(\lambda,E+i\eta,\theta,\omega) \in \mathbb{C}^\T $.
Since 
$\mu_{(0,0)}^{(E,\theta)} $ is supported on $ \Gamma(0,E+i0,\theta)$,
an
equivalent way of stating the conclusion \eqref{eq:weakconv} is
that the forward resolvents converge in distribution, i.e.,  for all $ x \in \mathbb{T} $ and almost all $ \theta \in \Xi $ 
\begin{equation}\label{eq:limitn}
        {\mathcal D}\!-\!\!\lim_{\substack{\lambda\to 0 \\ \eta \downarrow 0}} \;
        \Gamma_x(\lambda, \cdot +i \eta, \theta,\cdot) = \Gamma_x(0, \cdot +i0, \theta)
\end{equation} 
with $ \mathcal{D} := \mathcal{L}_I  \otimes \mathbb{P} $, cf.\
Definition~\ref{def:meas}  in Appendix~\ref{app:bounds}.
\item
        The family of measures $ \mu_{(\lambda, \eta)}^{(\theta)}$ is tight~\cite{MBauer,Bill68}.  This is readily deduced from:
           \begin{equation}\label{eq:Tfull}
           \inf_{t > 0 } \; \sup_{(\lambda,\eta) \in \mathbb{R} \times
[0,\infty)} \; \mu_{(\lambda, \eta)}^{(\theta)}\big(
           | \Gamma_x | > t \big)  = 0  \, ,
           \end{equation}
           for all $ x \in \mathbb{T} $ and $ \theta \in \Xi $, which follows from
Lemma~\ref{lemma:tight} in Appendix~\ref{app:bounds}.

\item The measures $ \mu_{(\lambda, \eta)}^{(\theta)} $ are of constant mass,
$|I|< \infty$.
For a family of such measures, tightness implies
that
   any sequence has (possibly many) weak accumulation
points~\cite{Bill68,MBauer}.  In order to prove the claimed convergence
\eqref{eq:weakconv} it suffices to show that any accumulation point of the given sequence 
coincides with $\mu_{(0,0)}^{(\theta)}$.
\end{enumerate}

We shall follow the path indicated by the last
observation.  By focusing on the  accumulation points  $ \mu^{(\theta)} $ of $
\mu_{(\lambda,\eta)}^{(\theta)} $, we may take advantage of the fact that in the
joint limit  $ \lambda , \eta \to 0 $  various
approximate statements which were outlined above take sharp form.
In particular, the recursion relation simplifies.  For the latter, we
shall make use of the following general principle.

\begin{proposition}\label{prop:eqconv}
Let $( \nu_\beta )_{\beta \in J} $ be a family of finite measures on
a polish space $ \Upsilon $, indexed by $\beta$ which
takes values in a topological space $J$.
Suppose that for each $\beta \in J$
\begin{equation} \label{eq:condition}
\varphi(\beta,Y) = 0 \quad \mbox{$\nu_\beta $-almost surely} \, ,
\end{equation}
with a  function
$ \varphi: J\times \Upsilon \to \C $ which:
\begin{itemize}
\item[i)] for every compact subset $ K \subset \Upsilon $ is equicontinuous
in $\beta$ over $ J\times K$, and
\item[ii)] at some $\beta_0\in J$ is continuous in $Y$, over $\Upsilon$.
\end{itemize}
Then, for each weak limit
$\nu =   \lim_{\beta \to \beta_0 } \nu_{\beta}$:
\begin{equation}
\varphi(\beta_0,Y) = 0 \quad \mbox{$\nu $-almost surely} \, .
\end{equation}
\end{proposition}
\begin{proof}   Since the space $\Upsilon$ is a  union of an
increasing family of compact sets,
it   suffices to show that for any compact set
        $ K  \subset \Upsilon $
\begin{equation}\label{eq:topr}
        \int_{K} \psi(\beta_0,Y) \,\nu(dY) = 0 \,
\end{equation}
with  $ \psi := | \varphi|/ (1+|\varphi|) $.
The integral in \eqref{eq:topr}
        may be rewritten as
        \begin{align}
         \int_{K} \!\psi(\beta_0,Y)\, \nu(dY) \leq & \int_K
\!\psi(\beta,Y) \, \nu_{\beta}(dY)
        + \sup_{Y \in K} \!\big|  \psi(\beta,Y) -  \psi(\beta_0,Y)
\big| \, \nu_\beta(K) \notag \\
        & + \Big| \int_K \psi(\beta_0,Y) \, \big( \nu(dY) -
\nu_{\beta}(dY) \big) \Big|.
\end{align}
Now, under the assumption \eqref{eq:condition} the first term
vanishes for every $ \beta $, and the conditions i) and ii) imply
that the second and  third term vanish in the limit $ \beta \to
\beta_0 $.
\qed\end{proof}

We now have the following characterization of
the possible accumulation points discussed above.

\begin{lemma}\label{lem:twop}
         Let  $ \mu^{(\theta)} $ be
        a  weak accumulation point for the family of measures
$\mu_{(\lambda, \eta)}^{(\theta)}$, with parameters $ (\lambda, \eta) $ in $ \R
\times (0,\infty) $ converging to $(0,0)$, i.e., 
        \begin{equation} \label{eq:weak}
        \lim_{\substack{\lambda\to 0 \\ \eta \downarrow 0}} \;
\mu_{(\lambda, \eta)}^{(\theta)} = \mu^{(\theta)} 
        \end{equation}
        Then: 
        \begin{enumerate}\itemsep1ex
        \item
        The limiting recursion relation
           \begin{equation}\label{eq:reclim}
             1 -  \Big(U_x(\theta)   - E - \sum_{y\in \mathcal{N}^+_x} \Gamma_y\Big) \,
\Gamma_x = 0
            \end{equation}
        holds  for all $ x \in \mathbb{T} $ and $ \mu^{(\theta)} $-almost all $ (E, \Gamma) $.
        \item For almost all $ (E,\theta) \in I \times \Xi $ the conditional
        measure $ \mu^{(E,\theta)} $\\[-1ex]
        \begin{enumerate}
        \item  satisfies the
        weak correlation condition~\eqref{ass:wc}.
        \item is supported on at most two points, i.e.,
        for all $ x     \in \T $  there exist $ I_x(E,\theta) \geq 0 $ and $ M_x(E,\theta) > 0 $ such that      
        for $ \mu^{(E,\theta)} $-almost all $ \Gamma $
        \begin{equation}\label{eq:ImMod}
           \Im \Gamma_x = I_x(E,\theta) \quad \mbox{and} \quad | \Gamma_x |
                =  M_x(E,\theta) \, .
        \end{equation}
     \end{enumerate} \end{enumerate}
\end{lemma}
\begin{proof}
        1. The first part is a consequence of
Proposition~\ref{prop:eqconv}.  To apply it, we let $\beta$ denote
the pair $(\lambda,\eta)$, with  $ J $
a neighborhood of $ \beta_0 := (0,0) $
                in $ \mathbb{R} \times (0,\infty) \cup \{(0,0)\} $. For
                $ \Upsilon $ we choose
        $ I \times \mathbb{C}^\T \times \R^\T $ endowed with the
product topology. The measures $ \nu_{(\lambda, \eta)} $
        are defined as the image of $ \mathcal{L}_I \otimes \mathbb{P} $ under the
        mapping
        \begin{equation}
         \big(E, \omega \big) \mapsto \big(E, \Gamma(\lambda,E+i\eta,\theta,\omega), V(\omega)  \big) \, ,
        \end{equation}
so that
        $ \mu_{(\lambda,\eta)}^{(\theta)} $ coincides with the projection of $\nu_{(\lambda, \eta)} $ 
        onto the first coordinates $ (E,\Gamma) $.
        Finally, we  set
        \begin{equation}
         \varphi\big((\lambda,\eta),E, \Gamma ,V \big) := 1 - 
        \Big( U_x(\theta) + \lambda \, V_x  - E - i \eta
        - \sum_{y\in \mathcal{N}^+_x} \Gamma_y \Big) \, \Gamma_x \, ,
        \end{equation}
so that the recursion relation \eqref{recur} can be expressed as
$\varphi\big((\lambda,\eta), E, \Gamma , V \big) = 0$.\\[0.5ex]
        2. We first note that \eqref{eq:weak} implies that there exists a set $ J \subseteq I $ of full Lebesgue measure
        and a subsequence $ \{(\lambda_k, \eta_k) \}_{k=0}^\infty $ of the original sequence such that
        \begin{equation}\label{eq:weak2}
                \lim_{k\to \infty} \;
                \mu_{(\lambda_k, \eta_k)}^{(E,\theta)} = \mu^{(E,\theta)} 
        \end{equation}
        for all $ E \in J $. \\[0.5ex]
        ~~~(a) The weak correlation property is inherited by weak limits,
        since it suffices to
        to verify \eqref{ass:wc} for bounded continuous functions
        $ F $ and $ G $. In that case
        the bound is implied by the continuity of the corresponding
        expectations under the weak convergence.\\[0.5ex]
        ~~~(b) We fix $ x \in \mathbb{T} $. Theorem~\ref{thm:cont} and
           Theorem~\ref{thm:imfoc} yield
           \begin{align}
                   \lim_{\substack{\lambda\to 0 \\ \eta \downarrow 0}}  \;
        \int_{I\times \Xi} \delta\big(\Im \Gamma_x(\lambda,E+i\eta,\theta,\cdot),\alpha
\big) \, dE \, p(d\theta) & = 0 \\
                    \lim_{\substack{\lambda\to 0 \\ \eta \downarrow 0}} \;
        \int_{I\times \Xi}
\delta\big(\big|\Gamma_x(\lambda,E+i\eta,\theta,\cdot)\big|^2,\alpha \big) \, dE\, p(d\theta) &
=0
           \end{align}
           for all $ \alpha \in (0,1/2] $.
        Lemma~\ref{lemma:seq2} in Appendix~\ref{App:error} and
\eqref{eq:weak2} thus imply that for
        Lebesgue-almost all $ E \in I $ both
        $ \Im \Gamma_x  $ and $ | \Gamma_x | $
        are $ \mu^{(E,\theta)} $-almost surely constant.
        Note that by \eqref{eq:reclim}, $ \Gamma_x \neq 0 $ $ \mu^{(E,\theta)}
$-almost surely.
\qed \end{proof}

Since a line and a circle can intersect in at most two points,
Eq.~\eqref{eq:ImMod} ensures that the $ \Gamma_x $-marginals
of $ \mu^{(E,\theta)} $  are supported on at most two points. In the next lemma,
we will actually prove that
these two points coincide.  Furthermore, using the uniqueness of periodic
solutions of the limiting recursion relation
(\ref{eq:reclim}) we shall conclude that this point coincides with
$ \Gamma_x(0,E+i0,\theta) $.  The chain of deductions can be presented as
follows.
\begin{lemma}\label{lemma:onep}
        Assume the situation of Lemma~\ref{lem:twop}. Then for
        almost all $ (E,\theta) \in I\times \Xi $ and all $ x \in \T $:
        \begin{enumerate}
        \item There exists $ \Phi_x(E,\theta) \in \C $ with $ \Im \Phi_x(E,\theta)
\geq 0 $ such that for $ \mu^{(E,\theta)} $-almost surely
        \begin{equation}
        \Gamma_x =  \Phi_x(E,\theta) \, .
        \end{equation}
        \item $ \Phi_x(E,\theta) = \Phi_0(E,\mathcal{S}^{|x|}\theta) $.
        \item $ \Phi_x(E,\theta) = \Gamma_x(0,E+i0,\theta) $.
        \end{enumerate}
\end{lemma}
\begin{proof}
        1. By \eqref{eq:ImMod} in Lemma~\ref{lem:twop} there exists
at most two points
        $ \Phi_x^\pm(E,\theta) \in \C \backslash \{0\} $
        with $ \Im \Phi_x^\pm(E,\theta) \geq 0 $ such that 
        $   \Gamma_x \in \{ \Phi_x^+(E,\theta),  \Phi_x^-(E,\theta) \} $ for $ \mu^{(E,\theta)} $-almost all $ \Gamma $.
        We will now prove by contradiction that these two points
coincide at every $ x \in \T $.

        Assume that there exists some $ y \in   \mathcal{N}_x^+ $ for
which   $\Phi_y^+(E,\theta) \neq \Phi_y^-(E,\theta)$.
        Since $ \mu^{(E,\theta)} $ is weakly correlated, we have
        \begin{equation}
           \mu^{(E,\theta)}\Bigg( \bigcap_{y \in
           \mathcal{N}_x^+} \Big( \Gamma_y = \Phi_y^{\pm}(E,\theta) \Big)\Bigg)
\geq \kappa\; \prod_{y \in \mathcal{N}_x^+}
           \mu^{(E,\theta)}\Big( \Gamma_y = \Phi_y^{\pm}(E,\theta) \Big) > 0,
         \end{equation}
        and similarly
        $ \mu^{(E,\theta)}\big( \Gamma_y = \Phi_y^+(E,\theta)  \  \mbox{and} \
\Gamma_{y'} = \Phi_y^-(E,\theta)
        \ \mbox{for all } y \neq y' \in \mathcal{N}_x^+  \big) >0 $,
which implies that the image measure
        of $
           \Gamma \mapsto U_x - E - \sum_{y \in \mathcal{N}_x^+} \Gamma_y
$
induced on $\mathbb{C}$ by $ \mu^{(E,\theta)} $ contains at least $ 3 $
points in its support. This is however not consistent with the
        limiting recursion relation (\ref{eq:reclim}) since the
measure induced on $\mathbb{C}$ by $ \Gamma_x^{-1}$,
        which is equal to one described above, is supported on only two points.\\[0.5ex]
        2. This property
        follows from the invariance of $ \mu_{(\lambda,\eta)}^{(E,\theta)} $
        under exchange of variables on disjoint forward subtrees and its covariance under radial shifts.\\[0.5ex]
         3. For almost all 
        $ E \in I $ the function $\theta \to \Phi_0(E,\theta) $ is for almost all $ \theta \in \Xi $ 
        a solution of the Schr\"odinger co-cycle~\eqref{eq:cocycle} with 
        $ \Im \Phi_0(E,\theta) \geq 0 $. It therefore coincides with  $ \Gamma_0(0,E+i0,\theta) $ by the uniqueness assumption 
        (which is verified in Appendix~\ref{app:proof} for the periodic case).  \qed
\end{proof}
We are now ready to conclude the
\begin{proof}[of Theorem~\ref{thm:meas}]
In order to prove the convergence asserted in \eqref{eq:weakconv},
it suffices to establish uniqueness of the accumulation point for the
measures $ \mu_{(\lambda,\eta)}^{(\theta)} $
with $\lambda\to 0  $ and $\eta \downarrow 0 $.
That was done in the argument which culminated in Lemma~\ref{lemma:onep},
which shows that for almost every $ \theta \in \Xi $ any such point coincides with  $ \mu_{(0,0)}^{(\theta)} $.
   \qed
\end{proof}

\section{Proof of the main result} 
The main results of this paper, Theorem~\ref{thm:main} and Theorem~\ref{thm:main2}, 
are a consequence of the convergence statements derived in the previous section. 
They culminate in the following  
\begin{theorem}\label{thm:main1} Let $ U(\theta) $ be radially stationary, assume that the co-cycle~\eqref{eq:cocycle} admits a unique solution
on $\Sigma_{ac}(0) $, and let 
$ V(\omega) $ satisfy {\bf A1}, {\bf A2} and {\bf
A3}. Moreover, let $  I \subseteq \Sigma_{\rm ac}(0) $
        be a bounded Borel set.
          Then for almost all $ \theta \in \Xi$:
          \begin{enumerate}\itemsep0.5ex
          \item The boundary values of the 
          forward resolvents of $ H_\lambda(\theta,\omega)  = T + U(\theta) + \lambda
V(\omega)  $ converge in distribution,
          \begin{equation}\label{eq:meas2}
                \mathcal{D}\!-\!\!\lim_{\lambda\to 0} \;
\Gamma_x(\lambda, \cdot + i0, \theta,\cdot) = \Gamma_x(0,\cdot +i0,\theta)
          \end{equation}
          for all $ x \in \mathbb{T} $, where $\mathcal{D} = \mathcal{L}_I\otimes\mathbb{P} $.
          \item
          If additionally $ I $ is closed and $\sigma_{\rm ac}(H_0(\theta)) \cap I = \sigma(H_0(\theta)) 
\cap I $, then
        the density of the ac component of the spectral measure
associated with the
          forward resolvents are $ L^1$-continuous at $ \lambda = 0 $
the sense that
           \begin{equation}\label{eq:main}
                \lim_{\lambda \to 0} \int_{I } \mathbb{E}\left[ \big|
\Im \Gamma_x(\lambda,E + i 0,\theta) - \Im \Gamma_x(0,E+i0,\theta) \big|
                  \right] \, dE = 0
        \end{equation}
        for all $ x \in \mathbb{T} $.
          \end{enumerate}
\end{theorem}
\begin{proof}
1.\ For a proof of (\ref{eq:meas2})
we let $ \varepsilon > 0 $ and use Fatou's lemma
\begin{align}\label{eq:Fatou}
   &\mathcal{L}_I\otimes\mathbb{P}\,\Big( \big|
    \Gamma_x(\lambda, \cdot + i 0, \theta,\cdot) - \Gamma_x(0,\cdot +i0,\theta) \big| >
    \varepsilon \Big)\notag \\
& \leq \liminf_{\eta \downarrow 0} \;
\mathcal{L}_I\otimes\mathbb{P}\, \Big(\big|
    \Gamma_x(\lambda, \cdot + i \eta, \theta,\cdot) - \Gamma_x(0,\cdot +i0,\theta) \big| >
    \varepsilon \Big) < \infty \, .
\end{align}
We now take the limit $ \lambda \to 0 $. Since \eqref{eq:limitn} holds
for any joint sequence of $ (\lambda,\eta) $
in $  \mathbb{R} \times (0,\infty) $, the right side of
\eqref{eq:Fatou} converges to zero in this limit.\\[0.5ex]
2.\ Eq.~\eqref{eq:main} follows from \eqref{eq:meas2} with the help
of Proposition~\ref{Prop:L1} in Appendix~\ref{app:bounds}.
To verify its assumptions we note that the resolvent
convergence \eqref{eq:srconv} implies
the weak convergence
$ \lim_{\lambda \to 0} \nu_\lambda^x(\theta,\omega) = \nu_0^x(\theta) $ of the
spectral measures associated with $ \Gamma_x $,
see~\eqref{eq:spectral} and \cite[Thm.~VIII.24]{RS:Vol1}. 
Moreover, since the spectrum of $ H_0(\theta) $ on $ I $ is purely ac, the spectral measures $ \nu_0^0(\theta) $ and hence 
$\nu_0^x(\theta) $ 
for all $ x \in \T $ are also purely ac on $ I $, cf.\ Appendix~\ref{app:proof}.  \qed
\end{proof}
The main result now follows from the special case $ x = 0 $ in the above the theorem.
\begin{proof}[of Theorem~\ref{thm:main} and Theorem~\ref{thm:main2}] 
The two statements can be proven simultaneously, as the only  difference in the argument is that in the general case of  Theorem~\ref{thm:main2} the uniqueness of solutions of \eqref{eq:cocycle}  is among  
the assumptions, whereas for the more specific case of Theorem~\ref{thm:main} this is established 
in Proposition~\ref{persol} of Appendix~\ref{app:proof}. \\[0.5ex]
1. By the non-randomness of the spectrum, it suffices to show that for any $ I \subseteq \Sigma_{\rm
ac}(0) $ and almost all $\theta $
\begin{equation}
         \lim_{\lambda \to 0} 
\mathbb{E}\left[\mathcal{L}_I\big( \sigma_{\rm
ac}(H_\lambda(\theta,\cdot)) \big)\right] = \mathcal{L}_I\big(I\big) \, .
\end{equation} 
We start the proof of this relation by observing that
\begin{align}
\mathcal{L}_I\big(I\big) & \geq
\mathbb{E}\left[\mathcal{L}_I\big( \sigma_{\rm
ac}(H_\lambda(\theta,\cdot)) \big)\right] \notag \\
& \geq \mathcal{L}_I\otimes \mathbb{P}\,\big\{
            0 < \Im \Gamma_0(\lambda, E+i0,\theta,\omega) < \infty \big\} \, .
\label{eq:lower1}
\end{align}
For any $ \varepsilon > 0 $ the set on the right side includes the collection of
$ (E,\omega)  $ for which $
\varepsilon < \Im \Gamma_0(0, E+i0,\theta)  < \infty $ and
$ \big| \Im \Gamma_0(\lambda, E+i0,\theta,\omega)  - \Im \Gamma_0(0, E+i0,\theta) \big| \leq
\varepsilon $. Accordingly, the right side of~\eqref{eq:lower1}
is bounded below by the difference of
\begin{equation}\label{eq:meas1}
            \mathcal{L}_I\otimes \mathbb{P}\,\big\{
            \big| \Im \Gamma_0(\lambda, E+i0,\theta,\omega)  - \Im \Gamma_0(0, E+i0,\theta) \big|
            \leq \varepsilon \big\}
\end{equation}
and
\begin{equation}\label{eq:meas3}
\mathcal{L}_I\big\{ \Im \Gamma_0(0, E+i0,\theta) \in [0,
\varepsilon] \cup \{\infty \} \big\}.
\end{equation}
As $ \lambda \to 0 $ the measure in \eqref{eq:meas1} converges to $
\mathcal{L}_I(I) $ for almost all $ \theta $ by Theorem~\ref{thm:main1}. Moreover,
as $ \varepsilon \downarrow 0 $ the measure in \eqref{eq:meas3}
converges to zero for almost all $ \theta $.\\[0.5ex]
2. This assertion coincides with \eqref{eq:main} in the special case $ x = 0 $. \qed
\end{proof} 


\section{Discussion}\label{sec:disc}

In what follows below, we include some additional comments on possible extensions of 
the results presented in this work.\\ 

\noindent 
1.  {\em The Green function at sites other than the root.}  
  Our discussion focused on the spectral measures associated 
with the vector $\delta_0 \in \ell^2(\T)$, the vector at the root.   
Theorem~\ref{thm:main} may, however, be extended to other sites on the tree graph, i.e., 
$\delta_0 $ may be replaced by $\delta_x $ in \eqref{eq:main1b}. 
For a proof, it is  useful to make the following observation. In order to study the Green function at $ x \neq 0 $ one may 
start from the collection of the forward resolvents of the ``siblings'' of $x$, i.e., $\{\Gamma_y\}_{|y|=|x|}$, 
which are then used as the input for a finite number of iterations of the suitably adjusted algebraic recursion relation~\eqref{recur}.     
Similarly, the statement can also be concluded for the fully symmetric tree, for which the root has $K+1$, rather than $K$ forward neighbors 
(whose forward trees are then of the kind discussed here).\\

\noindent 
2.  {\em Trees with non-constant branching.}  
The arguments presented here may easily be extended to  
trees where the branching numbers $ \{ K_x \}_{x \in \T} $ are non-constant but periodic functions of $ | x | $.  
For this case, the analysis applies with only the natural adjustments.\\ 

\noindent 
3.  {\em Almost periodic background operators.}  
The stability criterion expressed in Theorem~\ref{thm:main2} 
may be applied to radial operators with  quasi-periodic potentials,  
for example with $H_0(\theta)$ 
the almost-Mathieu operator~\cite{almostMatthew}.  
Such operators may exhibit both pure point and ac spectra in one dimension, and by Proposition~\ref{prop:1d} 
also on trees.  
The application of our analysis to such cases requires to verify the 
uniqueness of covariant solutions of the Schr\"odinger co-cycle 
\eqref{eq:cocycle}. This question is addressed in a subsequent publication \cite{AW}. \\

\noindent 
4.  {\em Location of the mobility edge.} 
We note that there is a gap between the results which address 
the possible location of pure point and ac spectrum for operators with weak disorder.   
For $H_{\lambda}(\omega) = T + V(\omega)$ the results of \cite{Klein98} as well as this work 
show that the location of the mobility edge for for $\lambda $ small is at  $|E|\ge  2\sqrt{K}$. 
Conversely, it is only known~\cite{A2} that the mobility edge approaches energies $|E|\le K+1$. The limiting value of the mobility edge is still unresolved.  The current guess is that it is the result of~\cite{A2} which needs to be improved (we thank Y.~Last for an illuminating discussion of this point).\\[1cm]


\appendix
\noindent {\bf  \Large  Appendix} 
\section{Schr\"odinger operators on tree graphs with radial
potentials}\label{app:proof} 
In this appendix we provide  some further details 
on the notions used in the paper and gather a few facts about 
the radial background operators considered in the main text.

\subsection{Radial Schr\"odinger operators}
The Schr\"odinger operator $ H = T + U $ on $ \ell^2(\T) $ is said to be
radial if the potential $ U $ is multiplication by a real-valued
function $\{ U_x \}_{x \in \T} $ which has the
radial symmetry of $ \T $, i.e.,
\begin{equation}
         U_x = U_y \qquad \mbox{for all $ x $, $ y \in \T $ with $ | x|
= | y | $}.
\end{equation}

The spectra of radial $ H$ are related to those of  the 
corresponding half-line operators. 

\begin{proposition} \label{prop:1d}
For any radial  $U$  the spectra of the operator 
$ H = T + U  $ on $ \ell^2(\T) $ and 
\begin{equation}
         H^+ = T+ K^{-1/2} \, U   \qquad \mbox{on $\ell^2(\N_0)$}
\end{equation}
satisfy
\begin{equation} \label{eq:spec}
        \sigma(H)    \supseteq    \sqrt{K} \; \sigma(H^+), \qquad
        \sigma_{\rm ac }(H)    =  \sqrt{K} \, \sigma_{\rm ac }(H^+).
\end{equation}
\end{proposition}
\begin{proof}
The action of $H$ on  $ \ell^2(\T) $ leaves invariant the subspace  
${\mathcal H}_{\rm rad}$
of  the radially symmetric functions.   
Under the partial isometry ${\mathcal U}: {\mathcal H}_{\rm rad} \to 
\ell^2(\N_0)$, defined by  $\big({\mathcal U} \, \psi\big)_x =  K^{|x|/2}\, \psi_{|x|}$,  
the restriction of $H$ to this subspace is unitarily equivalent to 
$ H^+ $.   
Hence  $ \sigma(H) \supseteq \sqrt{K} \, \sigma(H^+) $, and
analogously for the ac spectrum.

Under the action of $H$, $\delta_0$ is a cyclic vector 
for the symmetric subspace ${\mathcal H}_{\rm rad}$.  Therefore, 
to conclude the equality 
 $\sigma_{\rm ac }(H)    =  \sqrt{K} \, \sigma_{\rm ac }(H^+)$
it suffices to show that the support of the ac component of the spectral 
measure associated with an arbitrary site $x\in \T$ coincides with 
that of the root. 
The former is concentrated on the set of energies $ E \in \mathbb{R} $
for which
\begin{equation}\label{eq:ac}
        0 < \Im \big\langle \delta_x , \left( H - E -i0 \right)^{-1} 
\delta_x \big\rangle < \infty  \, , 
\end{equation}
and the latter on the set of energies where 
\begin{equation}\label{eq:ac2}
  0 < \Im \Gamma_{0}( E  + i 0 ) < \infty 
\end{equation} 

We claim that (\ref{eq:ac}) implies 
$  0 < \Im \Gamma_{y}( E  + i 0 ) < \infty \,$ for all $ y $ with $ |y | = | x | +1 $.
The reasoning involves  the radial symmetry, by which 
$\Gamma_x$ depends only on~$|x|$, and the recursion relation. 
Furthermore,  through the  recursion relation (\ref{eq:fpeq1}) below 
this leads to \eqref{eq:ac2}. 
\qed
\end{proof}

An alternative way to see the above unitary equivalence proceeds as follows.
Since $ U $ is radial, each forward resolvent $ \Gamma_x $
associated with $ H $ depends only on $ | x | $. 
With a slight abuse of notation, we will let $ \{\Gamma_n \}_{n \in
\mathbb{N}_0} $
stand for the values of the forward resolvents along one ray in the tree.
Using an analogous convention for $ U $, the recursion
relation~(\ref{recur}) can be rewritten as
\begin{equation}\label{eq:fpeq1}
        \Gamma_n = 
\frac{1}{U_n -z- K \Gamma_{n+1}}
        \qquad \mbox{for all $ n \in \mathbb{N}_0 $.}
\end{equation}
With suitable scaling, the recursion relation  \eqref{eq:fpeq1} is satisfied 
by  both the forward resolvent of $H$ and that of  $H^+$.  
As discussed in~\cite{FHS_04}, this relation has a unique solution for 
$\Im z>0$.

\subsection{Radially stationary Schr\"odinger operators}

A special case of radial Schr\"odinger operators 
are radially stationary ones with potential
$ U(\theta) $ defined in \eqref{eq:radstat} of Section~\ref{sec:radstat}.
The following proposition compiles some basic properties of such operators.

\begin{proposition}\label{prop:1d2}
For any radially stationary $ U(\theta) $ the ac spectrum of the 
Schr\"o\-dinger operator $ H(\theta) = T + U(\theta) $ on $ \ell^2(\T) $ has the properties:
\begin{enumerate}
\item It is related to the ac spectrum of 
\begin{equation}
         \widehat H(\theta) := T + K^{-1/2} \, U(\theta) \qquad \mbox{on $ \ell^2(\Z) $}
\end{equation}
for almost all $ \theta $,
        \begin{equation}\label{eq:1d}
                \sigma_{\rm ac}\big(H(\theta)\big) = \sqrt{K} \, \sigma_{\rm ac}\big(\widehat H(\theta)\big) \, .
        \end{equation}
\item It can be characterized for almost all $ \theta $ by
        \begin{equation}
                \sigma_{\rm ac}\big(H(\theta)\big) = \overline{\left\{ E \in \R \, : \, \gamma_0(E+i0) = 0 \right\}}^{\,\rm ess}        
        \end{equation}
        where $ \, \overline{(\cdot)}^{\, \rm ess} $ denotes the Lebesgue-essential closure and
        \begin{equation}
                \gamma_0(z) = \int_\Xi \log\left( \sqrt{K} \, \big| \Gamma_0(0,z,\theta) \big| \right) \, p(d\theta)
        \end{equation}
        stands for the Lyapunov exponent, cf.\ \eqref{def:Lya}.
\end{enumerate}
\end{proposition}
\begin{proof}
1. The first assertion is a consequence of Proposition~\ref{prop:1d} and Kotani theory \cite{Kot83,Sim83} which in particular 
        ensures that for almost all $ \theta $, $
         \sigma_{\rm ac}\big(H^+(\theta) \big) = \sigma_{\rm ac}\big(\widehat H(\theta) \big) $
        where $ H^+(\theta) $ acts on $ \ell^2(\N_0) $.\\[0.5ex]
2. The second assertion follows from \eqref{eq:1d} and again Kotani theory \cite{Kot83,Sim83}.
\qed
\end{proof}

Let us note that the disappearance  of the ac spectrum under arbitrarily weak random perturbations,
which was claimed in Proposition~\ref{prop:H0}, is implied by Proposition~\ref{prop:1d2} and well-known results 
about random Schr\"odinger operators in one dimension~\cite{CaLa90,PF}. \\     

\subsection{The radial periodic case}

The radially periodic potentials $ U $  fit into the framework of radially sationary ones: we choose the 
equidistribution $ p $ on  $ \tau $ integers $ \Xi = \{1,\dots,\tau\} $  on which $ \mathcal{S} \theta = (\theta+1) \mod \tau $ acts as
a periodic shift. \\

In this case the covariant form \eqref{eq:cocycle} of the recursion relation takes the form of a fixed point equation.
More precisely, introducing the family of 
M\"obius transformations $ \mathcal{T}(E,\theta)(\Gamma ) := \big( u(\theta) - E - K \Gamma \big)^{-1} $
and iterating over a period $ \tau $, we obtain
\begin{align}\label{eq:fpeq}
                \Gamma(\theta) & =  \mathcal{T}(E,\theta) \cdots
\mathcal{T}(E,\mathcal{S}^{\tau-1}\theta)\big(\Gamma(\theta)\big) \notag \\
        & =:
        \mathcal{S}(E,\theta)\big( \Gamma(\theta)\big) =: \frac{a(E,\theta) \, \Gamma(\theta)
        + b(E,\theta)}{c(E,\theta) \, \Gamma(\theta) +
        d(E,\theta)} \, .
\end{align}   
Here we used the fact that 
any  composition of real M\"obius transformations is a 
M\"obius transformations with some coefficients $ a(E,\theta) $, $
b(E,\theta) $, $c(E,\theta)$, $ d(E,\theta) \in \R $.

\begin{lemma}
For all, but a finite set of $ E \in \mathbb{R}$, the fixed point
equation \eqref{eq:fpeq} has
either two non-real, complex-conjugate solutions or two real solutions.
\end{lemma}
\begin{proof} The lemma is an immediate consequences of the following
observations:
\begin{itemize}
        \item[i)] the coefficients $ a(E,\theta) $, $ b(E,\theta) $, $c(E,\theta)$, $
d(E,\theta) $ are polynomials in $ E $ of degree at most $\tau$.
        \item[ii)] if $  c(E,\theta) \neq 0 $ for all  $ \theta \in \{1, \dots,
\tau\} $, then (\ref{eq:fpeq}) has one or
        two solutions in $ \mathbb{C} $, depending on the value of
        the discriminant 
        \begin{align}
         \varrho(E) & := \big( \tr  \mathcal{S}(E,\theta)\big)^2 - 4 \det \mathcal{S}(E,\theta) \notag \\
         & = \big( a(E,\theta) - d(E,\theta)\big)^2 - 4  b(E,\theta) c(E,\theta) \, ,
        \end{align}
         which does
        not depend on $ \theta $ for all $ E  $.
        \qed
\end{itemize}
\end{proof}

The following proposition collects some facts about radially 
periodic Schr\"o\-dinger operators. It includes a 
proof of the first statement in Proposition~\ref{prop:H0} and verifies that the co-cycle~\eqref{eq:cocycle} admits a unique solution
on $\Sigma_{ac}(0) $ in the periodic case.

\begin{proposition}\label{persol}
        Let  $ U $ be radially periodic. Then the ac spectrum of the operator $ H = T + U $ on $ \ell^2(\T) $
        is a union of intervals. In particular,
        \begin{equation} \label{eq:rhoneg}
          \sigma_{ac}( H) = \overline{ \{ E \in \mathbb{R} : \varrho(E) <
            0 \}},
        \end{equation}
        and therefore, for all but finitely many $E \in \sigma_{\rm ac}(H) $,
        the fixed point equation~\eqref{eq:fpeq} has two non-real,
        complex-conjugate solutions.
\end{proposition}
\begin{proof}
As was noted in Proposition \ref{prop:1d}, the ac spectrum of $H$ is concentrated
on those energies $E$ for which (\ref{eq:ac}), and therefore (\ref{eq:ac2}), holds.
For such energies, $\varrho(E) <0$. If $\varrho(E)<0$, it is clear
that $E \in \sigma_{\rm ac}(H)$. This proves (\ref{eq:rhoneg}). As
$\varrho$ is a polynomial in $E$, one has that the ac spectrum of
$H$ is a union of intervals, and away from those energies for which
$\varrho(E) = 0$, the fixed point equation \eqref{eq:fpeq} has two non-real,
complex-conjugate solutions.\qed
\end{proof}


\section{Some useful properties of resolvents}\label{app:bounds}
Let $ \nu $ be a finite Borel measure on $ \R $ and let $ F :
\mathbb{C}^+ \to \mathbb{C}^+ $ stand for its Stieltjes transform
given by
\begin{equation}\label{def:F}
        F(z) := \int_\R \frac{1}{t-z} \, \nu(dt).
\end{equation}
$ F $ is a Herglotz function, i.e.\ analytic and $ \Im F(z) > 0 $ for all
$  z \in \mathbb{C}^+ $.
In fact, every Herglotz function, which shares the property $
\sup_{\eta >0 } \eta \, \Im F(i\eta) < \infty $,
can be represented as a Stieltjes transform of a finite Borel measure 
\cite[App.~A]{PF}.
Examples of such functions are resolvents of self-adjoint
operators, in particular,
\begin{equation}\label{eq:spectral}
        \Gamma_x(\lambda,z,\theta,\omega) =  \int_\R \frac{1}{t-z} \;
\nu_\lambda^x(dt,\theta,\omega)
\end{equation}
where $ \nu_\lambda^x(\theta,\omega) $ is the spectral measure associated with
$ H^{\T_x}_\lambda(\theta,\omega) $ and $ \delta_x \in \ell^2(\T_x) $.\\

For the convenience of the reader, we collect some of basic
properties of Herglotz functions.
More information on this subject can be found in~\cite{Duren,Rudin}
or \cite[\S~7]{Kot85}, \cite{CaLa90}, and \cite[App.~A]{PF}.
\begin{proposition}\label{prop:Herglotz}
   Let $ F $ be the Stieltjes transform of a finite Borel measure $ \nu
$ on~$ \R $.
        \begin{enumerate}
        \item The boundary values $ F(E+i0 ) := \lim_{\eta \downarrow
0} F(E +i \eta) $ exist for almost every
        $ E \in \R $.
        \item The density of the ac component of $ \nu $ is given by
$ \,\Im F(E+i0) /\pi $.
        \item Let $ a $, $ b \in \R $ and $
s \in (0,1) $. Then
          \begin{equation}\label{eq:fracmom}
                  \int_a^b \left| F(E + i \eta) \right|^s \, dE \leq
                \frac{| b-a | + 2 (1-s)^{-1}}{\cos\big(\frac{\pi}{2}
s\big) }  \qquad \big[ =: B_s(a,b) \big]
          \end{equation}
        holds uniformly in $ \eta \geq  0 $ and $ \nu $.
        \end{enumerate}
\end{proposition}
\begin{proof}
Assertions 1. and 2. are taken from \cite[Thm.~2.2]{Duren} and
\cite[\S~7]{Kot85}.
In both cases the proof
uses the fact that $ F $ can be regarded as an analytic
function on the unit disk.

Assertion~3. borrows an idea from a theorem of Kolmogorov
\cite[Thm.~4.2]{Duren} and
is based on two observations concerning fractional moments:
          \begin{itemize}
          \item[i)]  If $\Im F \ge 0 $, then
           $
          | F|^s \leq \Re\left[ e^{-is\pi/2} \, F^s
\right]/\cos\big(\frac{\pi}{2} s \big).
          $
          \item[ii)] Let $\mathcal{C}_{a,b}$ be
          the rectangular contour joining  $a \to a +i \to b +i \to b   $, then
          $$
           \int_a^b  F(E+i\eta)^s \, dE  =
          \int_{\mathcal{C}_{a,b}}  F(z +i\eta)^s  \, dz.
          $$
          \end{itemize}
        The claim (\ref{eq:fracmom})  follows now, using the bound
        $|F(z +i\eta)|\leq 1/{\Im z} $.
\qed
\end{proof}

This paper mainly deals with Stieltjes transforms $ F(\cdot,\omega) $
of finite random Borel measures $ \nu(\omega) $ which depend measurably
on a parameter $ \omega $ from some probability space $ (\Omega, \mathbb{P}) $, see for example \eqref{eq:spectral}.
The following proposition establishes the equivalence of different
notions of convergence of the ac
spectral densities of such measures.
We take the opportunity to first recall
\begin{definition}\label{def:meas}
        A sequence of measurable functions $\left( f_n
\right)_{n=1}^\infty $ on $ I \times \Omega $ is said to
\emph{converge} to $ f $
        in \emph{$ \mathcal{L}_I \otimes \mathbb{P} $-measure} if
        \begin{equation}
                \lim_{n \to \infty} \mathcal{L}_I \otimes \mathbb{P}
\big( \left| f_n(E,\omega)
                        - f(E,\omega) \right| > \varepsilon \big) = 0
        \end{equation}
        for every $ \varepsilon > 0 $. We write: $  \mathcal{L}_I
\otimes \mathbb{P}-\!\lim_{n \to \infty} f_n = f $.
\end{definition}
It is fairly easy to see that $ L^1 $-convergence implies convergence
in measure \cite{MBauer}.
In the following we give a sufficient condition for the converse in
our special setting.
\begin{proposition}\label{Prop:L1}
Let $ ( F_n(\omega) )_{n = 1}^\infty $, $ F(\omega) $ be Stieltjes
transforms of finite random Borel measures
$ (\nu_n(\omega))_{n = 1}^\infty $, $ \nu(\omega) $ on $ \R $. Assume
that there exists
some closed Borel set $ I \subset \R $ such that for almost every $ \omega $
\begin{itemize}
        \item[i)] $ \displaystyle \; \lim_{n \to \infty}
\nu_{n}(\omega) = \nu(\omega) $ in the sense of weak convergence on $
I $, and
        \item[ii)] $ \displaystyle \; \nu(\omega) $ is purely ac on $ I $.
\end{itemize}
Then the convergence of the ac densities in
$  \mathcal{L}_I \otimes\mathbb{P} $-measure,
\begin{equation}\label{eq:convmeas}
        \mathcal{L}_I\otimes\mathbb{P}-\!\!\lim_{n\to \infty} \;  \Im
F_n(\,\cdot+i0,\cdot) = \Im F(\,\cdot+i0,\cdot),
\end{equation}
implies their $ L^1 $-convergence
\begin{equation}\label{eq:L1}
          \lim_{n \to \infty} \int_{I} \mathbb{E}\left[ \big| \Im
F_n(E+i0,\cdot) - \Im F(E+i0,\cdot) \big| \right] dE= 0.
\end{equation}
\end{proposition}
\begin{proof}
        Every subsequence of $ ( \Im F_n )_{n = 1}^\infty $
          has a subsequence $ \big(\!\Im F_{n_k}\big)_{k=1}^\infty $ for which
          \begin{equation}\label{eq:pointconvF}
          \lim_{k \to \infty} \Im F_{n_k}(E +i 0, \omega) = \Im F(E +i 
0, \omega)
          \end{equation}
          for $ \mathcal{L}_I \otimes \mathbb{P} $-almost all $ (E,
\omega) \in I \times \Omega $. This statement is in fact equivalent
          \cite[Thm.~20.7]{MBauer} to the convergence in measure
\eqref{eq:convmeas}.
        Moreover,
        \begin{align}\label{eq:boundlimsup}
        \limsup_{n \to \infty} \int_I \mathbb{E}\left[ \Im F_n(E+i0,\cdot)
\right] dE \leq & \; \pi \, \limsup_{n \to \infty}
        \, \mathbb{E}\left[\nu_n\big(I\big)\right]
        =  \pi \, \mathbb{E}\left[\nu\big(I\big)\right] \notag \\
        =
        & \int_I \mathbb{E}\left[ \Im F(E+i0,\cdot) \right] dE.
        \end{align}
        Here the first equality is a consequence of i) and ii), cf.\ 
\cite[Thm.~30.12]{MBauer}.
The last equality 
in~\eqref{eq:boundlimsup}  expresses the fact that by assumption 
ii) the measure $\nu(\omega)$ has no singular component in $I$.  
Eq.~(\ref{eq:L1}) now follows for subsequences fulfilling
\eqref{eq:pointconvF} from Fatou's lemma with remainder \cite{LL01}.
This completes the proof since every subsequence has a
subsequence for which (\ref{eq:L1}) holds. \qed
\end{proof}

In the remainder of this appendix we collect more specialized bounds
on the forward resolvents \eqref{eq:spectral}.
Our first estimate concerns the rareness of large values
these resolvent.
\begin{lemma}\label{lemma:tight}
        Let $ a $, $ b\in  \R $ and $ t > 0 $. Moreover, let
        $ x \in \mathbb{T}$ and $ \lambda \in \R $. Then for all $
\eta \geq 0 $:
        \begin{equation}\label{eq:tight}
            \int_a^b\mathbb{P}\big(|\Gamma_x(\lambda, E + i \eta) | > t
\big) \,   dE   \leq   \frac{ B_s(a,b)}{t^s},
          \end{equation}
          where $ B_s(a,b) $ was defined in \eqref{eq:fracmom}.
\end{lemma}
\begin{proof}
This is a consequence of the Chebychev-Markov
inequality and \eqref{eq:fracmom}. \qed
\end{proof}

Our last estimate will guarantee the finiteness of logarithmic
moments of the forward resolvents \eqref{eq:spectral}.
\begin{lemma}\label{lem:Gup}
    Let $ \lambda \in \mathbb{R} $, and $ z \in \mathbb{C}^+ $. Then
          for every $ x \in \mathbb{T} $,
     \begin{align}\label{Gup}
      \big| \log \Gamma_x(\lambda, z,\omega) \big|
       \leq    2 \, \log\big(1+ K \, (\Im z)^{-1} \big) + \log( 1 + | z
|) &\notag\\
        + \log( 1 + | U_x| + | \lambda|\,  | V_x|)  + \pi. &
    \end{align}
\end{lemma}
\begin{proof}
    The proof is based on the fact that for any $ \Gamma\in
\mathbb{C}^+ $ one has
    \begin{equation}\label{eq:log}
      | \log  \Gamma | \leq \left| \log | \Gamma | \right| + \pi \leq
    \log^+ | \Gamma |  + \log^+ | \Gamma|^{-1} + \pi,
    \end{equation}
    where
    $ \log^+ x := \max\{ 0 , \log x \} $ denotes the positive part of the
    logarithm of $ x > 0 $.
    Inserting $ \Gamma_x $, the first term is bounded according to
    $ \log^+ |\Gamma_x| \leq \log^+ (\Im z)^{-1} \leq \log\big(1+ K \,
(\Im z)^{-1} \big) $
    for all $ \lambda \in \mathbb{R} $ and $ z \in \mathbb{C}^+ $.
    To bound the second term we employ
    the recursion relation (\ref{recur}) and the triangle inequality to obtain
    \begin{align}
        \log^+ |\Gamma_x |^{-1}
       & = \log^+ \Big|z - U_x- \lambda V_x 
       - \sum_{y\in \mathcal{N}_x^{+}} \Gamma_y \Big| \notag \\
       & \leq  \log^+ \left( |U_x| +  | \lambda|\, |  V_x| + |  z | + K
(\Im z)^{-1}\right)
    \end{align}
    for all $ \lambda \in \mathbb{R} $ and $ z \in \mathbb{C}^+ $.
\qed\end{proof}

\section{Extension of a result by Kotani and Simon}\label{App:thouless}
For one-dimensional random Schr\"odinger operators, it is known
\cite{CaLa90,PF} that the ac spectrum can be characterized by the
essential
closure of the set of energies for which the Lyapunov exponent
vanishes. This result has two parts. Ishii \cite{Ishi73} and Pastur
\cite{Pas80,PF} showed that the positivity of the Lyapunov exponent
implies the vanishing of the ac spectrum. Their result does
\emph{not} extend to trees as is illustrated
          by the following two examples:\\

Suppose $ U = 0 $ and recall from \eqref{eq:Tfree} that the spectrum of $ T $ on $
\ell^2(\T) $ is purely ac.
          \begin{enumerate}
          \item If $ \{ \omega_x \}_{x \in \mathbb{T}} $ are iid Cauchy 
variables with
          $$
           \mathbb{P}\left( \omega_x \in A \right) = \frac{1}{\pi} \int_A
\frac{dv }{v^2 + \sigma^2} \quad
                  \mbox{with some $ \sigma > 0 $ and all Borel $ A
\subset \mathbb{R} $},
          $$
          then Theorem~\ref{thm:main} shows that for sufficiently small
$ \lambda > 0 $ and almost surely the
          spectrum of $ H_\lambda(\omega) = T + \lambda V(\omega) $ has
an ac component within its spectrum.
          \item If $ V(\omega) $ is \emph{radial}, i.e., given by $
V_x(\omega) = \omega_{|x|} $, where $ \{ \omega_{n} \}_{n \in \mathbb{N}_0} $ are
iid
          Cauchy variables as in 1, then by Proposition~\ref{prop:H0},  for all
          $ \lambda > 0 $ and almost every $ \omega $ the
          spectrum of $ H_\lambda(\omega) = T + \lambda V(\omega) $ has
\emph{no} ac component.
          \end{enumerate}

          In both cases, the Lyapunov exponents are identical. In fact,
a contour-integration argument shows that
          \begin{equation}
           \gamma_\lambda(E)
          =  - \log \left(\sqrt{K} \, \big| \Gamma_0(0, E + i \lambda
\sigma ) \big|\right) > 0
          \end{equation}
          for every $ E \in \sigma(T)  $ and all $ \lambda > 0 $.
        Here the positivity follows from
          the explicit expression for
          $ \Gamma_0(0,z) $.\\

The converse of the Ishii-Pastur result is due to Kotani
\cite{Kot83}, who showed that for
one-dimensional, continuum random Schr\"odinger operators, zero
Lyapunov exponent on some Borel set of positive measure implies the
existence of ac spectrum there.
Shortly thereafter, Simon demonstrated that an
analogous result holds in the discrete setting \cite{Sim83}.
This result partially extends to trees.
\begin{theorem}\label{thm:KotSim} Let $U(\theta)$ be radially stationary and
$ V(\omega) $ satisfy {\bf A1}--{\bf A3}. Let $ \lambda \in \R $
and
    suppose $I \subset \mathbb{R}$ is an open interval for which
    $\gamma_{\lambda}(E+i0) = 0 $ for almost every $E \in
    I$. Then, $\Sigma_{ac}( \lambda ) \cap I \neq
\emptyset \, $.
\end{theorem}
\begin{proof} We use a slight variation of the argument in \cite{Sim83}.
Since $ \Re w_\lambda(E+i0) = 0 $ for almost all $ E \in I $, the
Schwartz-reflection principle \cite{Rudin} (see also
\cite[Lemma~7.5]{Kot85}) ensures that
the function $ w_\lambda $ defined in (\ref{def:w}) has an analytic
continuation through $ I $
in the lower half-plane.
In particular, this implies that its derivative exists through $ E
\in I $ and one has
\begin{equation}
          \lim_{\eta \downarrow 0} \frac{\gamma_\lambda(E+i\eta)}{\eta}
= \lim_{\eta \downarrow 0} \frac{d \gamma_\lambda(E+i\eta)}{d \eta} <
          \infty.
\end{equation}
Using Fatou and Lemma~\ref{lem:kotani} below, one sees that
\begin{align}
          &\int_\Xi \mathbb{E}\left[ \left(\Im
\Gamma_{0}(\lambda, E + i0,\theta,\cdot)\right)^{-1} \right]\, p(d\theta)
\notag \\
          & \leq
          \liminf_{\eta \downarrow 0} \int_\Xi  \mathbb{E}\left[
\left(\Im \Gamma_{0}(\lambda, E + i\eta,\theta,\cdot)+
      \frac{\eta}{2  K } \right)^{-1}\right]\, p(d\theta) < \infty,
\end{align}
and therefore, $ \Im \Gamma_0(\lambda, E + i0,\theta,\omega) > 0 $ for
almost every $ E \in I $ and $ \theta,\omega  $.
\qed
\end{proof}
The proof of the previous theorem relied on the following
\begin{lemma} \label{lem:kotani}
Let $ \lambda \in \mathbb{R} $ and $z \in \mathbb{C}^+$, then
\begin{equation} \label{eq:lyb}
        \int_\Xi \mathbb{E} \left[ \left( \Im  \Gamma_{0} ( \lambda, z,\theta,\cdot) +
      \frac{\Im z}{2  K }\right)^{-1} \right] \, p(d\theta) \leq  2 K
\, \frac{\gamma_{\lambda}(z)}{\Im z }.
\end{equation}
\end{lemma}

\begin{proof}
Again our proof is similar to Simon's argument \cite{Sim83}. Using
(\ref{imrecur}) and the Jensen inequality we obtain
\begin{eqnarray} \label{eq:gbd}
\log \left( \frac{ \Im \Gamma_x}{ K \,
|\Gamma_x|^2} \right) & = & \log \left( \frac{ \Im z}{K} +
\frac{1}{K} \sum_{y \in \mathcal{N}_x^{+}} \Im
\Gamma_y  \right)  \nonumber \\ & \geq & \frac{1}{K}
\sum_{y \in \mathcal{N}_x^{+}} \log \left( \frac{ \Im z}{
    K} + \Im \Gamma_y \right).
\end{eqnarray}
Taking the probabilistic expectation and using the fact that all
forward neighbors are
identically distributed, the right side is equal to
$ \mathbb{E} \log \big(
   \Im z/ K + \Im \Gamma_y  \big) $ for any $ y \in \mathcal{N}_x^{+} $.
Averaging over $ \theta $, we thus obtain the estimate
\begin{eqnarray}
2 \, \gamma_{\lambda}(z) &\geq &   \int_\Xi   \mathbb{E} \left[
\log \left( 1 +
        \frac{ \Im z}{K \, \Im \Gamma_{0}(\lambda, z,\theta)} \right)
    \right]\, p(d\theta) \nonumber \\
    & \geq &   \int_\Xi   \mathbb{E} \left[ \frac{ 2 \Im z}{2 K \,
\Im \Gamma_{0}(\lambda, z,\theta) + \Im z} \right]\, p(d\theta)\, ,
\end{eqnarray}
where the last inequality follows from the fact that $\log(1+x)\geq
\frac{2x}{2+ x}$ for any $x \geq 0$. \qed
\end{proof}

\section{On relative errors}\label{App:error}

Throughout this appendix
let $ (\Omega, \mathbb{P} ) $ be some probability space and
$ X  $  stand
for a positive random variable.
Our main object of interest is the relative width $ \delta(X, \cdot)
$ associated with the distribution of $ X $, which was introduced in
Definition~\ref{def:relwitdth}.\\

The following lemma collects some useful rules of
calculus associated with addition, multiplication and division of
positive random variables.
\begin{lemma}\label{lemma:delta}
          Let $ X $ and  $ (X_j)_{j=1}^K $ be a collection of positive
random variables. Let $ \alpha $ and $ (\alpha_j)_{j=1}^K $ be
          a collection of numbers with values in $ (0,1/2] $. Then:
          \begin{enumerate}\itemsep0.5ex
          \item $ \;\delta(X,\alpha_1) \leq \delta(X,\alpha_2) \quad $
if $ \alpha_1 \geq \alpha_2 $.
          \item   $\; \delta(\eta + X, \alpha) \leq \delta(X,\alpha)
\quad$ if $ \eta \geq 0 $.
          \item   $\; \delta(1/X,\alpha) = \delta(X,\alpha) $.
          \item   $\; \delta\left(\prod_{j=1}^K X_j , \sum_{j=1}^K
\alpha_j \right) \leq
                    \sum_{j=1}^K \delta(X_j, \alpha_j) \quad$ if $
\sum_{j= 1}^K \alpha_j \in (0,1/2] $
          \item   $ \;\delta\left( \sum_{j=1}^K X_j , \alpha K \right)
\leq \delta(X,\alpha) \quad$
          if $ (X_j)_{j=1}^K $ are identically distributed and $ K
\alpha \leq 1/2 $.
          \end{enumerate}
\end{lemma}
\begin{proof}
          Assertion~1 follows from the fact that $ \alpha  \mapsto
\xi_\mp(X, \alpha) $ is monotone increasing/decreasing
          respectively.
          A proof of assertion~2 is based on
          the identity
          $ \xi_\pm(\eta + X, \alpha) = \eta + \xi_\pm(X, \alpha) $ which yields
                  \begin{equation}
                          \delta(\eta + X, \alpha) =
                          1 - \frac{\eta + \xi_-(X, \alpha)}{\eta +
\xi_+(X, \alpha)}
                          \leq \delta(X,\alpha)
                  \end{equation}
          by  monotonicity in $ \eta \geq 0 $. Assertion~3 follows from
the equality $ \xi_\mp(1/X, \alpha) = 1/ \xi_\pm(X, \alpha) $.
          For a proof of assertion~4 we estimate
          \begin{align}
          & \mathbb{P}\left( \prod_{j=1}^K X_j < \prod_{j=1}^K
\xi_-(X_j, \alpha_j) \right)\notag \\
          & \leq
          \mathbb{P}\Big( \mbox{There is some $ j $ with}
                     \; X_j < \xi_-(X_j, \alpha_j) \Big)
           \leq \sum_{j=1}^K \alpha_j
          \end{align}
          and hence $\xi_-\left( \prod_{j=1}^K X_j , \sum_{j=1}^K \alpha_j \right) \geq
                  \prod_{j=1}^K \xi_-(X_j, \alpha_j) $.
          The same lines of reasoning also yield the bound
          $ \xi_+\left( \prod_{j=1}^K X_j , \sum_{j=1}^K \alpha_j\right) \leq
          \prod_{j=1}^K \xi_+(X_j, \alpha_j) $ such that
          \begin{align}
                  \delta\left(\prod_{j=1}^K X_j , \sum_{j=1}^K \alpha_j \right)
                  & \leq 1 - \prod_{j=1}^K \frac{\xi_-(X_j,
\alpha_j)}{\xi_+(X_j, \alpha_j)} \notag \\
                  & \leq \sum_{j=1}^K \left( 1 -
                          \frac{\xi_-(X_j, \alpha_j)}{\xi_+(X_j,
\alpha_j)} \right)
                  = \sum_{j=1}^K \delta(X_j, \alpha_j) .
          \end{align}
          The last inequality follows by induction on $ K $.
          For a proof of assertion~5 we estimate
          \begin{align}
                  \mathbb{P}\Bigg(\sum_{j=1}^K X_j < \sum_{j=1}^K
\xi_-(X_j, \alpha) \Bigg)
                  & \leq
                  \mathbb{P}\Big( \mbox{There is some $ j $ with} \;
                          X_j < \xi_-(X_j, \alpha) \Big) \notag \\
                  & \leq \alpha K
          \end{align}
          and hence $ \xi_-\left(\sum_{j=1}^K X_j , \alpha K \right) \geq
          \sum_{j=1}^K \xi_-(X_j, \alpha) = K \xi_-(X_1, \alpha) $, because
          the random variables $ (X_j) $ are identically distributed.
Similarly, we obtain
          the bound $ \xi_+\big(\sum_{j=1}^K X_j , \alpha K \big) \leq
K \xi_+(X_1, \alpha) $
          which proves the claim.
\qed\end{proof}

The next lemma employs the relative width of a single random variable
as a lower bound to certain expectation values
involving two identically distributed random variables under a weak
correlation assumption.
\begin{lemma}\label{lemma:wcorr}
          Let $ X_1 $ and $ X_2 $ be identically distributed positive
random variables.
 Suppose that there exists a constant $ \kappa \in (0, 1]$ 
such that
          \begin{equation}
                   \mathbb{P}\left( X_1 \in A_1 \; \mbox{and} \; X_2
\in A_2 \right)
                          \geq \kappa \; \mathbb{P}\left( X_1 \in A_1
\right) \mathbb{P}\left( X_2 \in A_2 \right)
          \end{equation}
          for all pairs of Borel sets $ A_1, A_2 \subset (0, \infty) $. Then
          \begin{equation}\label{eq:lemma3}
                  \mathbb{E}\left[ \left(\frac{X_1 - X_2}{X_1 +
X_2}\right)^2\right] \geq \frac{\kappa}{2}
                  \big[ \alpha \, \delta(X_1 , \alpha ) \big]^2.
          \end{equation}
          for all $ \alpha \in (0,1/2] $.
\end{lemma}
\begin{proof}
          In the event that $ X_1 \le \xi_-(X_1,\alpha) $ and $ X_2 \ge
\xi_+(X_1, \alpha) $, one has
          \begin{equation}
                   \left|\frac{X_1 - X_2}{X_1 + X_2}\right|
                  \geq \frac{\xi_+(X_1, \alpha) -
\xi_-(X_1,\alpha)}{\xi_+(X_1, \alpha) + \xi_-(X_1,\alpha) }
                  = \frac{\delta(X_1 , \alpha )}{2 - \delta(X_1 , \alpha )}
                   \geq \frac{\delta(X_1 , \alpha )}{2}.
          \end{equation}
The same holds true if $ X_2 \le \xi_-(X_1,\alpha) $ and 
$ X_1 \ge \xi_+(X_1, \alpha) $.
Therefore the left side in (\ref{eq:lemma3}) is 
bounded from below by $ \delta(X_1 , \alpha )^2 / 4 $ times
          \begin{multline}
                          \mathbb{P}\big( X_1 \le \xi_-(X_1,\alpha) \;
\mbox{and} \; X_2 \ge \xi_+(X_1, \alpha) \big) \\
                          +
                          \mathbb{P}\big( X_2 \le \xi_-(X_1,\alpha) \;
\mbox{and} \; X_1 \ge \xi_+(X_1, \alpha) \big) \\
                          \geq 2 \kappa\, \mathbb{P}\big( X_1 \le
\xi_-(X_1,\alpha) \big) \; \mathbb{P}\big(X_1 \ge \xi_+(X_1, \alpha)
\big)
                          \geq 2 \kappa\alpha^2.
          \end{multline}
\qed\end{proof}

Rather elementary considerations yield the following  useful  statement.  
We note that $ \xi_\pm(\nu,\alpha) $ are defined as in \eqref{eq:defxi} for any measure $ \nu $ on $ [0,\infty) $.

\begin{lemma}\label{lemma:seq}
    Let $ \{\nu_n\}_{n= 1}^\infty $ be a sequence of  probability measures on 
$(0,\infty)$, which has a weak limit: $  \lim_{n \to \infty} \nu_n = \nu $.  Then, 
for each $\alpha \in (0,1/2] $ 
\begin{equation}\label{eq:xibounds}
\limsup_{n\to \infty}  \xi_-(\nu_n,\alpha) \leq \xi_-(\nu,\alpha)\leq \xi_+(\nu,\alpha) \le \liminf_{n\to \infty}  \xi_+(\nu_n,\alpha) 
\end{equation}
Furthermore, if  for each 
$\alpha \in (0,1/2] \cap \mathbb{Q}$
\begin{equation}
 \liminf_{n \to \infty} \delta (\nu_n , \alpha) =0,
\end{equation}  
then $\nu$ is supported on 
a single point, i.e., there is $\xi\in \R$ such that $\nu \{\xi\} =1$. 
\end{lemma} 
\begin{proof} 
By its definition, $\xi_+(\nu,\alpha)$ is the smallest real number $\xi \in \R $ such that 
   $\nu[\xi-\varepsilon, \infty)  > \alpha$  for any $\varepsilon >0$, and it suffices to restrict here the attention to $\varepsilon $ for which  
$\nu \{ \xi - \varepsilon \} =0$.  For such, we may deduce that for large enough $n$ also: 
$\nu_n[\xi-\varepsilon, \infty)  > \alpha$, and hence $\xi_+(\nu_n,\alpha) \ge \xi - \varepsilon$.   From this we conclude that 
$ \xi_+(\nu,\alpha) \le \liminf_{n\to \infty}  \xi_+(\nu_n,\alpha) $
and, by analogous reasoning,  
$\xi_-(\nu,\alpha) \ge \limsup_{n\to \infty}  \xi_-(\nu_n,\alpha)$.   

For a proof of the second assertion we distinguish two cases. If $ \xi_+(\nu,\alpha) = 0 $ for some $ \alpha \in  (0,1/2] $,
 then $ \nu\{0\} = 1 $, because $ \nu $ is supported on $ [0,\infty ) $. Otherwise, if $ \xi_+(\nu,\alpha) >  0 $ 
for all $ \alpha \in  (0,1/2] $, then $ \delta (\nu, \alpha) = 1 -  \xi_-(\nu,\alpha)/ \xi_+(\nu,\alpha) $ is well-defined and 
\eqref{eq:xibounds} implies
\begin{equation}\label{eq:deltanull}
      \delta (\nu, \alpha) \ \le  \ \liminf_{n \to \infty} \delta(\nu_n, \alpha)  \, .
    \end{equation} 
The claims now readily follow as, in this case, zero relative width implies zero absolute width.
\qed 
\end{proof} 

In the main text, we need the following consequence of the preceding lemma. 

\begin{lemma}\label{lemma:seq2}
 Let $ (T,\rho) $ be a finite measure space and $ \{\nu_n^t\}_{n= 1}^\infty $ be a measurable 
family of sequences of  probability measures on 
$(0,\infty)$ which are indexed by $ t \in T $. Assume that $  \lim_{n \to \infty} \nu_n^t = \nu^t $ as a weak limit for 
almost all $ t \in T $.     
Suppose further that
\begin{equation}\label{eq:intdelta}
        \lim_{n \to \infty} \int_T \delta( \nu_n^t, \alpha) \, \rho(dt) = 0
\end{equation}
for all $ \alpha \in (0,1/2] $.
Then for almost all $ t \in T $ there exists $ \xi^t \in
[0,\infty) $ such that $ \nu^t $ is supported on~$ \xi^t $.
\end{lemma}
\begin{proof}
  From \eqref{eq:intdelta} we conclude that there exists some set
$ S \subseteq T $ of full $ \rho $ measure and
some subsequence $ \{ \nu_{n_k}^t\}_{k= 1}^\infty $ such that
\begin{equation}
        \lim_{k \to \infty} \delta(\nu_{n_k}^t, \alpha) = 0
\end{equation}
for all $ t \in S $ and all $ \alpha \in  (0,1/2] \cap \mathbb{Q} $.
Since the weak convergence also holds down the subsequence, i.e., 
$ \lim_{k \to \infty} \nu^t_{n_k} = \nu^t $ at  almost 
every $t$, the claim follows from Lemma~\ref{lemma:seq}. \qed
\end{proof}


\begin{acknowledgement}

We are much indebted to Thomas Chen for useful comments.  
MA thanks Uzy Smilansky and the Weizmann Institute for gracious hospitality.  This work was supported by the Einstein Center for Theoretical Physics and the Minerva Center for Nonlinear Physics at the Weizmann Institute, by the US National Science Foundation, and by the Deutsche Forschungsgemeinschaft. 

\end{acknowledgement} %


\begin{thebibliography}{10}

\bibitem{Abou73}
R.~{Abou-Chacra}, P.~W. 
Anderson, and D.~J. Thouless.
\newblock A selfconsistent theory of 
localization.
\newblock {\em J. Phys. C: Solid State Phys.}, 
6:1734--1752, 1973.

\bibitem{Abou74}
R.~Abou-Chacra and D.~J. 
Thouless.
\newblock Self-consistent theory of localization. {II}. 
localization near the
  band edges.
\newblock {\em J. Phys. C: Solid 
State Phys.}, 7:65--75, 1974.

\bibitem{AM}
M.~Aizenman and 
S.~Molchanov.
\newblock Localization at large disorder and at extreme 
energies: an elementary
  derivation.
\newblock {\em Commun. Math. 
Phys.}, 157:245, 1993.

\bibitem{A2}
M.~Aizenman.
\newblock 
Localization at weak disorder: some elementary bounds.
\newblock {\em 
Rev. Math. Phys.}, 6:1163--1182, 1994.

\bibitem{AW}
M.~Aizenman and S. Warzel.
\newblock 
Persistence under weak disorder of AC spectra of quasi-periodic Schr\"odinger operators on tree graphs.
\newblock 
Preprint math-ph/0504084.


\bibitem{Anderson}
P.~W. 
Anderson.
\newblock Absence of diffusion in certain random 
lattices.
\newblock {\em Phys. Rev.}, 109:1492--1505, 
1958.

\bibitem{MBauer}
H.~Bauer.
\newblock {\em Measure and 
integration theory}.
\newblock de Gruyter, Berlin, 
2001.

\bibitem{Bill68}
P.~Billingsley.
\newblock {\em Convergence of 
probability measures}.
\newblock Wiley, New York, 
1968.

\bibitem{CaLa90}
R.~Carmona and J.~Lacroix.
\newblock {\em 
Spectral theory of random {S}chr\"odinger operators}.
\newblock 
Birkh\"auser, Boston, 1990.

\bibitem{Duren}
P.~L. Duren.
\newblock 
{\em Theory of $ H^p $ spaces}.
\newblock Academic, New York, 
1970.

\bibitem{FHS_04}
R. Froese, D. Hasler and W. Spitzer.
\newblock 
{\em Transfer matrices, hyperbolic geometry and absolutely 
continuous spectrum for some discrete Schr\"odinger 
operators on graphs}.  
\newblock 
Preprint, 2004.

\bibitem{GoMo77}
{I. Ya.} Goldsheid, S.~Molchanov, and 
L.~Pastur.
\newblock A pure point spectrum of the stochastic 
one-dimensional schr\"odinger
  operator.
\newblock {\em Funct. Anal. 
Appl.}, 11:1--8, 1977.

\bibitem{HLMW01}
T.~Hupfer, H.~Leschke, P.~M\"uller and S.~Warzel.
\newblock
Existence and uniqueness of the integrated density of states for
Schr{\"o}dinger operators
with magnetic fields and unbounded random potentials.
\newblock
{\em Rev. Math. Phys.} 13:1547--1581, 2001.


\bibitem{Ishi73}
K.~Ishii.
\newblock 
Localization of eigenstates and transport phenomena in the one-dimensional disordered system.
\newblock {\em Supp. Progr. Theor. 
Phys.}, 53:77--138, 1973.


\bibitem{almostMatthew}
S.~Ya.~Jitormirskaya.
\newblock 
Metal-insulator transition for the almost-Mathieu operator
\newblock {\em Ann. Math.}, 150:1159--1175, 1999.

\bibitem{Klein95}
A.~Klein.
\newblock {The 
Anderson metal-insulator transition on the Bethe lattice.}
\newblock 
In D.~Iagolnitzer, editor, {\em Proceedings of the XIth 
international
  congress on mathematical physics, Paris, France, July 
18-23, 1994}, pages
  383--391. International Press, Cambridge, MA, 
1995.

\bibitem{Klein96}
A.~Klein.
\newblock {Spreading of wave 
packets in the Anderson model on the Bethe
  lattice}.
\newblock {\em 
Commun. Math. Phys.}, 177:755--773, 
1996.

\bibitem{Klein98}
A.~Klein.
\newblock {Extended states in the 
Anderson model on the Bethe lattice}.
\newblock {\em Adv. Math.}, 
133:163--184, 1998.

\bibitem{Kot83}
S.~Kotani.
\newblock Ljapunov 
indices determine absolute continuous spectra of stationary
  one 
dimensional {S}chr{\"o}dinger operators.
\newblock In K.~Ito, editor, 
{\em Proc. Taneguchi Itern. Symp. on Stochastic
  Ananlysis}, pages 
225--247, Amsterdam, 1983. North Holland.

\bibitem{Kot85}
S.~Kotani.
\newblock 
One-dimensional random {S}chr{\"o}dinger operators and {H}erglotz
functions.
\newblock In K.~Ito, editor, {\em Taneguchi Symp. PMMP}, 
pages 219--250,
  Amsterdam, 1985. North 
Holland.


\bibitem{LL01}
E. H.~Lieb ans M. Loss.
\newblock
{\em Analysis}.
\newblock
      $ 2 $nd edition. Amer. Math. Soc., Providence, RI, 2001.

\bibitem{MilDer93}
J.~D. Miller and B.~Derrida.
\newblock 
Weak disorder expansion for the {A}nderson model on a tree.
\newblock 
{\em J. Stat. Phys}, 75:357--388, 1993.

\bibitem{MiFy91}
A.~D. Mirlin and Y.~V. Fyodorov.
\newblock Localization transition in the 
{A}nderson model on the {B}ethe
  lattice: spontaneous symmetry 
breaking and correlation functions.
\newblock {\em Nucl. Phys. B}, 
366:507--532, 1991.

\bibitem{PF}
L.~Pastur and A.~Figotin.
\newblock 
{\em Spectra of random and almost-periodic operators}.
\newblock 
Springer-Verlag, Berlin, 1992.

\bibitem{Pas80}
L.~A. 
Pastur.
\newblock Spectral properties of disordered systems in the 
one body
  approximation.
\newblock {\em Commun. Math. Phys.}, 
75:167--196, 1980.

\bibitem{RS:Vol1}
M.~Reed and B.~Simon.
\newblock 
{\em Methods of modern mathematical physics {I}: {F}unctional analysis}.
\newblock Academic Press Inc., New York, second edition, 
1980.

\bibitem{Rudin}
W.~Rudin.
\newblock {\em Real and complex 
analysis}.
\newblock McGraw-Hill, New York, third edition, 
1987.

\bibitem{Sim83}
B.~Simon.
\newblock Kotani theory for 
one-dimensional {J}acobi matrices.
\newblock {\em Commun. Math. 
Phys.}, 89:227--234, 1983.

\bibitem{Stoll}
P.~Stollmann.
\newblock 
{\em Caught by disorder: bound states in random media}.
\newblock 
Birkh\"auser, Boston, 2001.

\end{thebibliography}
%
%

\end{document}